%% file: template.tex
\documentclass[]{article}

\usepackage{arxiv}

\usepackage[utf8]{inputenc} 
\usepackage[T1]{fontenc}    
\usepackage{hyperref}       
\usepackage{url}            
\usepackage{booktabs}       
\usepackage{amsfonts}       
\usepackage{nicefrac}       
\usepackage{microtype}      
\usepackage{lipsum}
\usepackage{graphicx}
\usepackage{tikz}
\usepackage{tabularx} 
\usepackage{subcaption}
\usepackage{svg}            
\usepackage{layouts}        
\usepackage{comment}        
\usepackage{siunitx}

\title{A nation-wide experiment: fuel tax cuts and almost free public transport for three months in Germany - Report 2  First wave results}

\author{
Fabienne Cantner\\
Technical University of Munich\\
TUM School of Management\\
TUMCS for Biotechnology \& Sustainability \\ 
Am Essigberg 3, 94315 Straubing\\
\texttt{fabienne.cantner@tum.de}\\
\And
Nico Nachtigall\\
Technical University of Munich\\
TUM School of Engineering and Design\\
Chair of Automotive Technology \\
Boltzmannstrasse 15, 85748 Garching\\
\texttt{nico.nachtigall@tum.de}\\
\And
Lisa S. Hamm \\
Technical University of Munich \\
TUM School of Engineering and Design\\
Chair of Traffic Engineering and Control\\
Arcisstrasse 21, 80333 Munich \\
\texttt{lisa.hamm@tum.de} \\
\And
Andrea Cadavid Isaza\\
Technical University of Munich\\
TUM School of Engineering and Design\\
Chair of Renewable and sustainable energy systems\\
Lichtenbergstraße 4a, 85748 Garching\\
\texttt{andrea.cadavid@tum.de}\\
\And
Lennart Adenaw\\
Technical University of Munich\\
TUM School of Engineering and Design\\
Chair of Automotive Technology \\
Boltzmannstrasse 15, 85748 Garching\\
\texttt{lennart.adenaw@tum.de}\\
\And 
Allister Loder \\
Technical University of Munich\\
TUM School of Engineering and Design\\
Chair of Traffic Engineering and Control\\
Arcisstrasse 21, 80333 Munich \\
\texttt{allister.loder@tum.de}\\
\And
Markus B. Siewert\\
Munich School of Politics and Public Policy\\
TUM Think Tank\\
Richard-Wagner-Straße 1, 80333 München\\
\texttt{markus.siewert@hfp.tum.de}
\And
Sebastian Goerg\\
Technical University of Munich\\
TUM School of Management\\
TUMCS for Biotechnology \& Sustainability \\ 
Am Essigberg 3, 94315 Straubing\\
\texttt{sebastian.goerg@tum.de}\\
\And
Markus Lienkamp\\
Technical University of Munich\\
TUM School of Engineering and Design\\
Chair of Automotive Technology \\
Boltzmannstrasse 15, 85748 Garching\\
\texttt{lienkamp@tum.de}\\
\And
Klaus Bogenberger\\
Technical University of Munich\\
TUM School of Engineering and Design\\
Chair of Traffic Engineering and Control\\
Arcisstrasse 21, 80333 Munich\\
\texttt{klaus.bogenberger@tum.de}\\
}

\begin{document}
\maketitle
\begin{abstract}
In spring 2022, the German federal government agreed on a set of measures that aim at reducing households' financial burden resulting from a recent price increase, especially in energy and mobility. These measures include among others, a nation-wide public transport ticket for 9\ EUR per month and a fuel tax cut that reduces fuel prices by more than 15\,\%. In transportation research this is an almost unprecedented behavioral experiment. It allows to study not only behavioral responses in mode choice and induced demand but also to assess the effectiveness of transport policy instruments. We observe this natural experiment with a three-wave survey and an app-based travel diary on a sample of hundreds of participants as well as an analysis of traffic counts. In this second report, we update the information on study participation, provide first insights on the smartphone app usage as well as insights on the first wave results, particularly on the 9\ EUR-ticket purchase intention.
\end{abstract}


\section{Introduction}


In transportation research, it is quite unlikely to observe or even perform real-world experiments in terms of travel behavior or traffic flow. There are few notable exceptions: subway strikes suddenly make one important alternative mode not available anymore  \cite{Anderson2014,Adler2016}, a global pandemic changes travelers' preferences for traveling at all or traveling collectively with others \cite{Molloy2021}, or a bridge collapse forces travelers to alter their daily activities \cite{Zhu2010}. However, in 2022 the German federal government announced in response to a sharp increase in energy and consumer prices a set of measures that partially offset the cost increases for households. Among these are a public transport ticket at 9\ EUR per month\footnote{\url{https://www.bundesregierung.de/breg-de/aktuelles/9-euro-ticket-2028756}} for traveling all across Germany in public transport, except for long-distance train services (e.g., ICE, TGV, night trains), as well as a tax cut on gasoline and diesel, resulting in a cost reduction of about 15\ \% for car drivers\footnote{\url{https://www.bundesfinanzministerium.de/Content/DE/Standardartikel/Themen/Schlaglichter/Entlastungen/schnelle-spuerbare-entlastungen.html}}. Both measures are limited to three months, namely June, July and August 2022. As of mid June, more than 16\ million tickets have been sold\footnote{
\url{https://www.tagesschau.de/wirtschaft/unternehmen/neun-euro-ticket-135.html}
}, while it seems that the fuel tax cut did not reach consumers due to generally increased fuel prices and oil companies are accused of not forwarding the tax cuts to consumers \footnote{\url{https://apnews.com/article/politics-business-germany-prices-deb85a000d63cd57b76446d9c90c3e18}}. 

For the Munich metropolitan region, Germany, we designed a study comprising three elements. The three elements are: (i) a three-wave survey before, during and after the introduction of cost-saving measures; (ii) a smartphone app based measurement of travel behavior and activities during the same period; (iii) an analysis of aggregated traffic counts and mobility indicators. We will use data from 2019 (pre-COVID-19) and data from shortly before the cost reduction measures as the control group. In addition, the three-wave survey is presented to a German representative sample. The main goal of the study is to investigate the effectiveness of the cost-saving measures with focus on the behavioral impact of the 9\ EUR-ticket on mode choice \cite{ben1985discrete}, rebound effects \cite{Greening2000,Hymel2010}, and induced demand \cite{Weis2009}. Further details on the study design can be found in our first report \cite{reportone}.

In this second report, we first provide an update on the study participation in Section \ref{sec:participation}; second, we present first insights into the mobility app usage in Section \ref{sec:app}; third, Section \ref{sec:survey} summarizes first results of the first-wave survey; fourth, in Section \ref{sec:ticket} we show preliminary results regarding the intentions to buy and use the 9\ EUR-ticket. 

\section{Participation update} \label{sec:participation}

\input{_sections/participation}

\section{App insights} \label{sec:app}

\input{_sections/app}

\section{First wave insights} \label{sec:survey}

\subsection{Socio-economic characteristics} \label{sec:socio}

\input{_sections/socio}

\subsection{Travel behavior} \label{sec:mobility}

\input{_sections/mobility}

\subsection{Energy and households expenditures} \label{sec:energy}

\input{_sections/energy}

\section{Insights 9\ EUR-Ticket intentions} \label{sec:ticket}

\input{_sections/Ticket}

\section{Discussion and outlook}

In this second report, we have provided first insights into the first-wave survey of our study. We have seen that our nation-wide sample can be considered representative when compared to the MiD travel survey, while our Munich-oriented sample most likely requires a re-weighting as our sample has too many participants with academic degrees and bicycle users. We have further seen that many respondents already experience the cost increases caused by the sharp increase in energy prices. Thus, it is perhaps not surprising that many respondents favor the introduction of the 9 EUR ticket as a cost reduction measure. Importantly, we see that personal opinions and the financial burden the cost increases factor into the overall support for the ticket and the intention for buying the ticket; here we have to observe in the second and third wave how these factors develop.

The next report, no. 03, will provide further insights into the travel behavior obtained through the app. In particular, how the data is processed and how the mobility indicators are being calculated.

\section*{Acknowledgements}

The authors would like to thank the TUM Think Tank at the Munich School of Politics and Public Policy led by Urs Gasser for their financial and organizational support and the TUM Board of Management for supporting personally the genesis of the project. The authors thank the company MOTIONTAG for their efforts in producing the app at unprecedented speed. Further, the authors would like thank everyone who supported us in recruiting participants, especially Oliver May-Beckmann and Ulrich Meyer from M Cube and TUM, respectively.

\bibliographystyle{unsrt}  
\bibliography{references}  



\end{document}

%% file: _sections/participation.tex
In total, 2\ 258 individuals registered for our study. 921 participated in the representative sample and 1337 in the Munich-oriented sample. In the latter case, 260 registered only for the survey and 1\ 075 registered for the app and the survey; 2 individuals prematurely opted out of the study.

\begin{figure}
     \centering
     \begin{subfigure}[b]{0.45\textwidth}
         \centering
         \includegraphics[width=\textwidth]{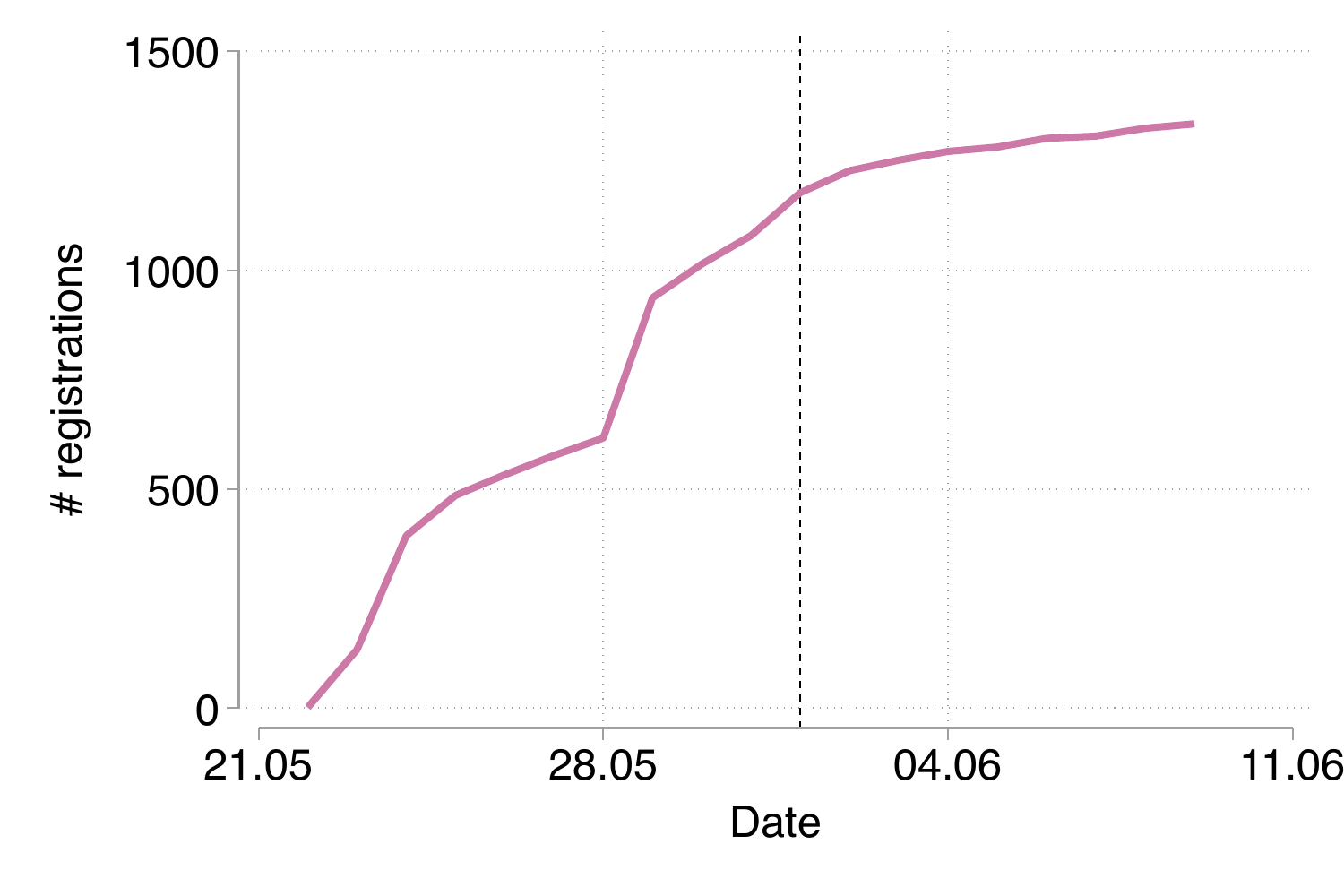}
         \caption{Registrations.}
         \label{fig:registration}
     \end{subfigure}
     \hfill
     \begin{subfigure}[b]{0.45\textwidth}
         \centering
         \includegraphics[width=\textwidth]{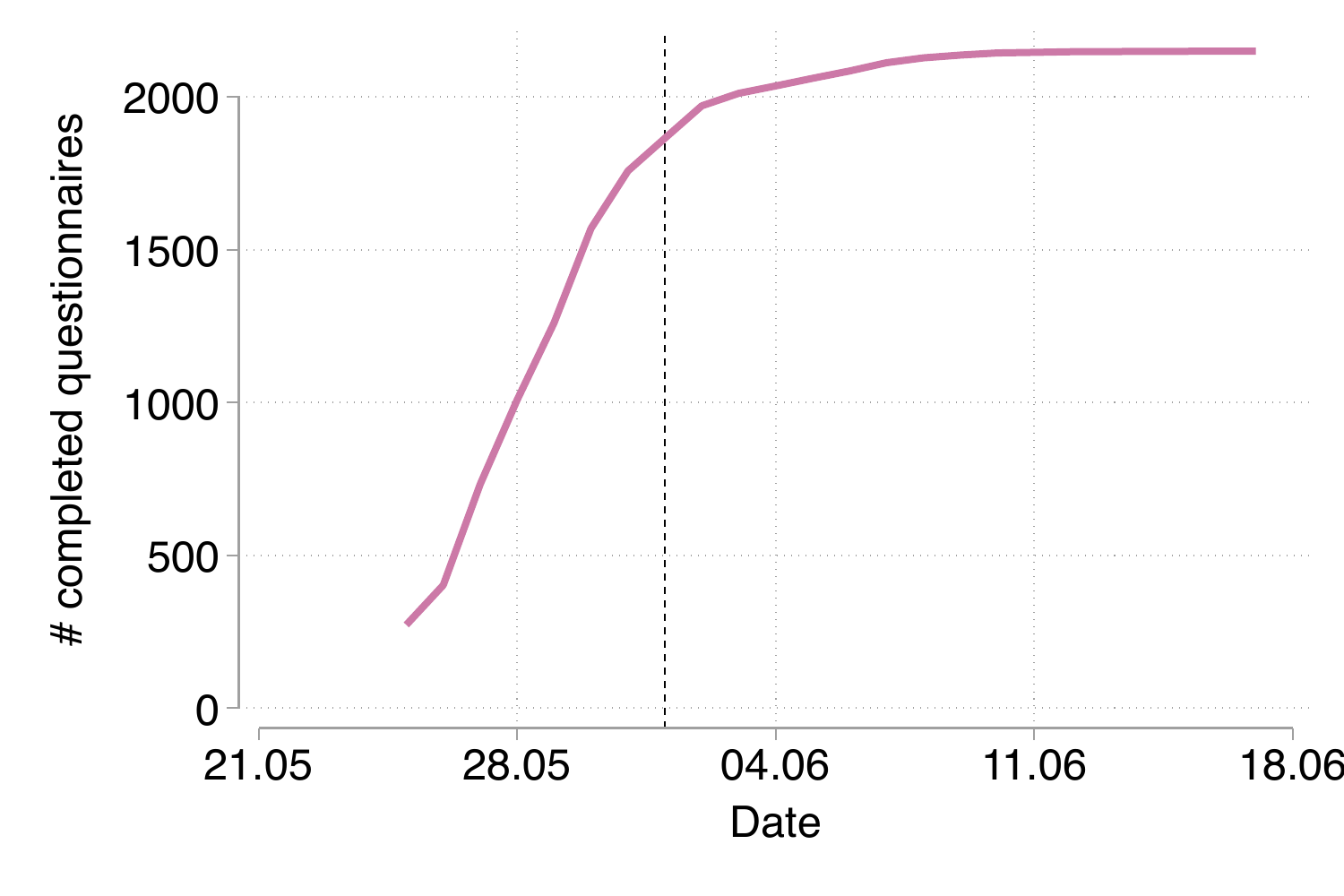}
         \caption{Survey completion.}
         \label{fig:survey}
     \end{subfigure}
     
     \begin{subfigure}[b]{0.45\textwidth}
         \centering
         \includegraphics[width=\textwidth]{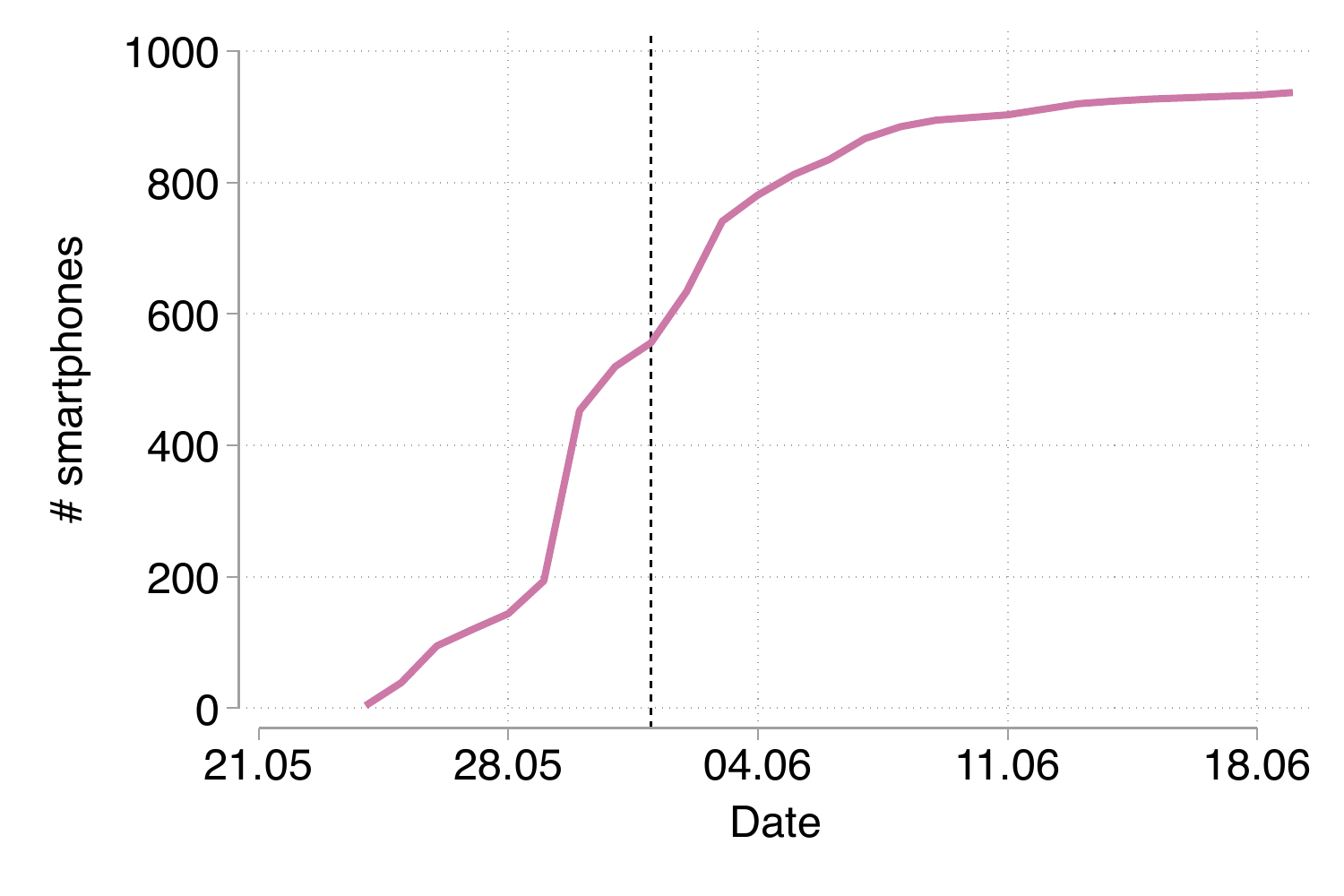}
         \caption{App activation.}
         \label{fig:app}
     \end{subfigure}
        \caption{Participation status.}
        \label{fig:participation}
\end{figure}

Figure \ref{fig:participation} shows the time-series of the participation status as of June 18, 2022. In Figure \ref{fig:registration} we see that for the Munich-oriented sample roughly 80\ \% of the sample have registered before the start of the cost reduction measures. The two sharp increases in registrations result from a press conference and a TV broadcast. The registration was closed on June 9, 2022 because the cost reduction measures have already started and obtaining unbiased before survey responses is getting less likely. 

Figure \ref{fig:survey} shows the development of the first wave's completion numbers of the pooled sample, \mbox{i.e.}, the \mbox{Munich-oriented} and representative sample. Before the introduction of the cost reduction measures on June 1, 2022, 98\ \% of the representative sample, 60\ \% of the \mbox{Munich-oriented} \mbox{survey-only} sample, and 65\ \% of the \mbox{Munich-oriented} \mbox{app-and-survey} sample finished the first survey. The first wave survey was closed on June 16, 2022. In the end, we have a survey completion rate of 100\ \% for the representative sample. In the Munich-oriented sample we have a survey completion rate of 87\ \% for the \mbox{only-survey} group and 92\ \% in the \mbox{app-and-survey} group. 

The app activation pattern shown in \mbox{Figure \ref{fig:app}} is primarily governed by the availability of both smartphone apps in the respective app stores. The iOS app is available since May 25, 2022 and the Android app is available since May 30, 2022. The app has been activated on 521 smartphones before the start of the cost reduction measures. As of June 17, 2022 the app has been activated on more than 900 smartphones. 

%% file: _sections/app.tex

This section deals with the smartphone application used to record the participants' mobility behavior during the course of the study. It sheds light on both, the structure and mechanics of the application and the usage intensity and behavior shown by participants from the beginning of the tracking experiment until June 11, 2022, our current evaluation date.

The application itself is provided for both large smartphone operating systems: Android and iOS. To make the installation as easy as possible, the app is directly available through the Android Play Store and the iOS App Store, respectively. The app is called "Mobilität.Leben" and is available within both application distribution platforms until this experiment concludes. Every registered participant receives a personal activation code via email. As soon as a participant has downloaded the app, this registration code can be used to activate the app and identify the user. Then, the mobility tracking commences. The app runs as a silent background task and does not expect any user input to start, stop or resume tracking. However, participants can decide to deactivate or pause tracking at any time and they can partially correct the collected data. Since the app relies on global positioning services, location services must be switched on and disruptions or disturbances of the location services signal will skew the recorded waypoints. The recorded data is transferred to a server backend which \mbox{pre-processes} the raw data to extract two \mbox{so-called} storyline elements from the raw waypoints: a \textit{stay}, which indicates activities, \mbox{i.e.} phases of no or locally restricted movement not attributable to travel; and a \textit{track}, which represents movement caused by transportation. For each stay, a presumed activity type and duration is inferred from the raw data. For each track, a mode of transport and a travel distance is estimated. The participants are able to see their own data once all storyline elements of a trip have been processed by the backend; they can decide to correct any storyline element manually. 

\begin{figure}
	\centering
	\includegraphics[width=\textwidth]{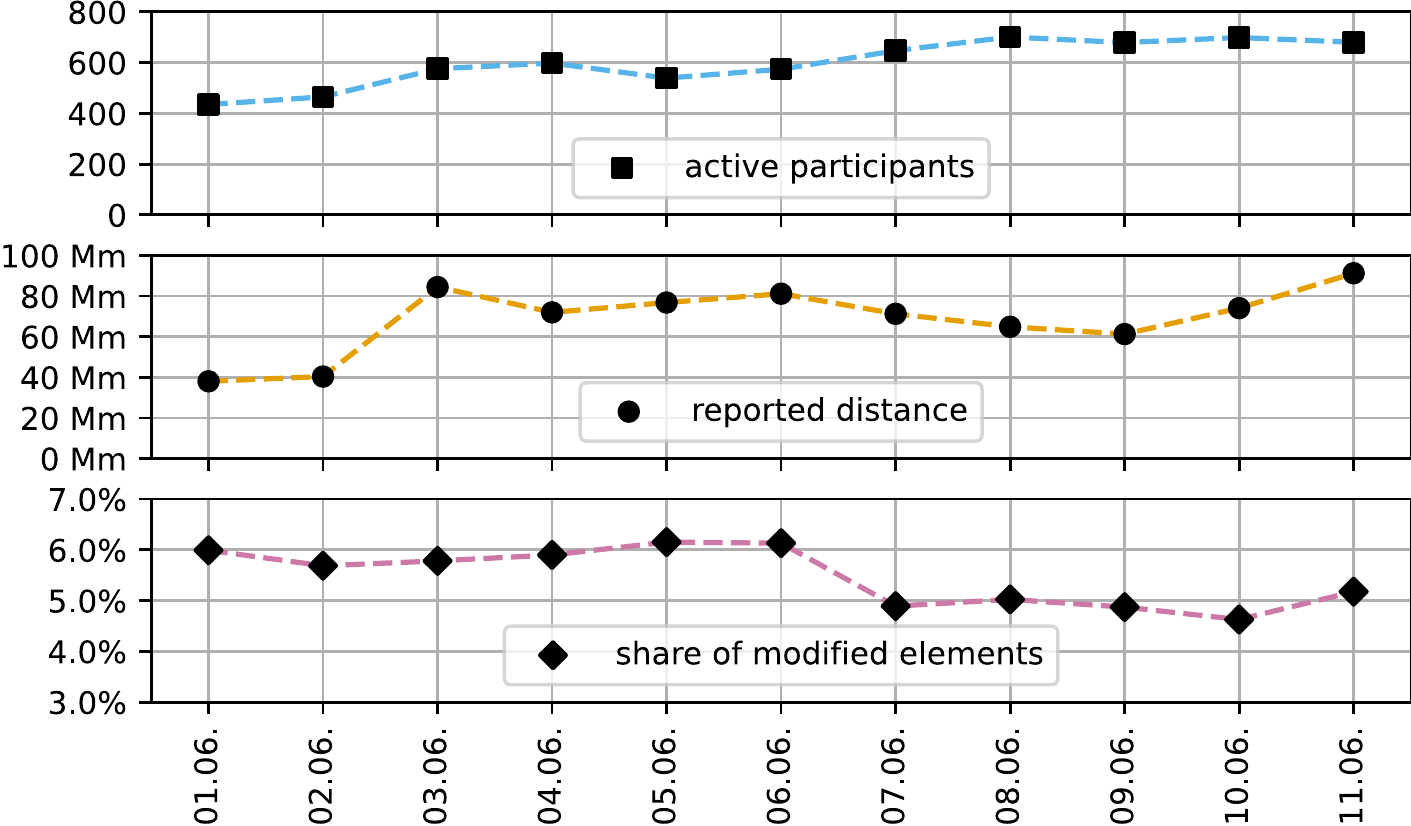}
	\caption{App usage indicators; from top to bottom: number of participants who recorded at least one trip per day, total recorded distance per day, and share of manually modified storyline elements per day}
	\label{fig:app_usage}
\end{figure}

Between June 1, 2022 and June 11, 2022, 885 participants have started recording their travel behavior with our app, \mbox{i.e.,} contributed at least with a single data point. \mbox{Figure \ref{fig:app_usage}} provides an overview of three different app usage indicators. The upper part of the figure reveals the number of users who recorded at least one trip or activity for each day of the experiment. The figure shows that this number has grown from the beginning of the tracking experiment with a little more of 400 active daily users to constantly over 600 contributing users at any given day after June 6, 2022. The growing number of active daily users is coherent with the app activations indicated by \mbox{Figure \ref{fig:participation}}. The slightly decrease of active users on Sunday June 5, 2022 is inline with results reported in \mbox{\cite{dlr125879}}: more people do not leave home during Sundays in comparison to the other weekdays.
On each day of the ongoing study, between \SI{38029}{\kilo\meter} and \SI{91250}{\kilo\meter} of travel were recorded by participants. The large span between these figures can again be explained by the growing number of active participants during the first week of the study. Additionally, fluctuations can be expected due to weekends and public holidays. A reliant daily total distance is yet to be established especially since this reporting period of 10 days comprises two Saturdays, one Sunday, and Pentecost. Probably also related to the school holidays in Bavaria are the high recorded distances with in average \SI{107}{\kilo\meter} per active user and day. When excluding trips of a length of more than \SI{200}{\kilo\meter} the average traveled distance is with \SI{66}{\kilo\meter} per active user and day significantly lower, suggesting that \mbox{vacation-related} travel activities have a large impact on the data collected so far. The possibility to adjust tracks or stays manually has constantly been utilized by less than 10\% of the participants. The willingness to manually correct storyline elements appears to be decreasing; whether this is an indication of a disengagement from the participants' side or of an improved backend estimation of storyline elements will be a matter of investigation in further reports.   

Summarizing the presented first insights into app usage, it can be said that a profound analysis of mobility behavior for the sample is not yet possible because of the dynamics of a study \mbox{ramp-up} as well as a possible interference of the  many public and school holidays in the first days of the study. Nevertheless, the app mechanics described in this section will remain an important factor in the subsequent analysis of the collected tracking data; a first look at the number of active users and recorded daily distances is promising with regard to the usage intensity among participants. The data base of manually adjusted tracks and stays may provide valuable insights into the data sets later on as these entries offer validation and error estimation potentials for the whole sample.

%% file: _sections/socio.tex
\begin{figure}
\centering
    \includegraphics[width=11cm]{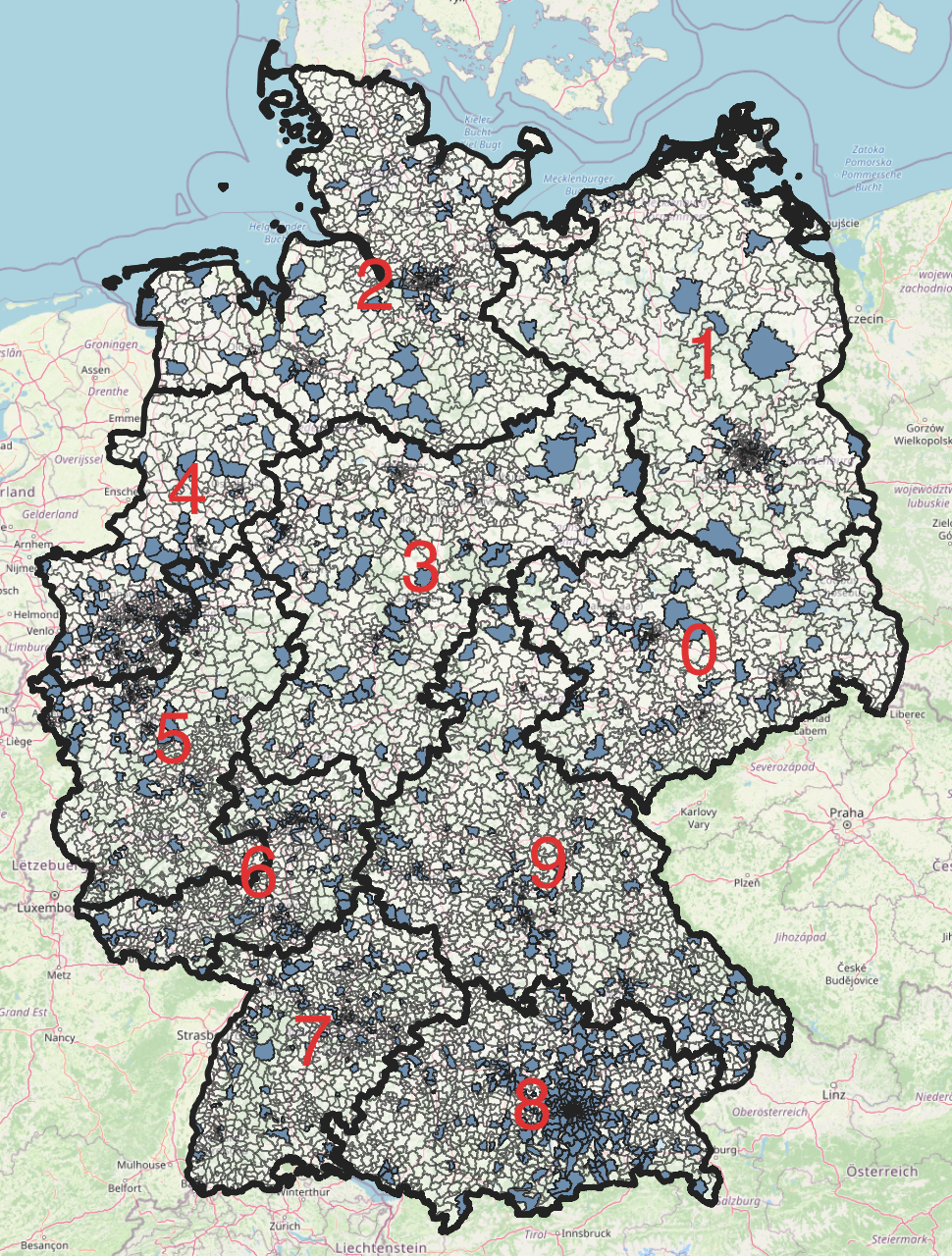}
    \caption{Home locations of the pooled sample. Zones highlighted in blue have at least one respondent in the pooled sample. The postal code area boundaries are emphasized and the respective area number is given in red.}
    \label{fig:homelocations}
\end{figure}

The pooled sample is distributed across Germany as seen in Figure \ref{fig:homelocations} that shows if a postal code area is considered in our study with at least one participant.Table \ref{tab:1} shows the distribution of the nation-wide and Munich-oriented sample across postal code areas in Germany; it contrasts these numbers with the distribution of the overall population in Germany. More than half of the respondents (55.5\ \%) in the pooled sample reside in postal area 8 that covers the Munich metropolitan area around Munich including Ulm, Landshut, Augusburg and Ingolstadt. This oversampling clearly results from the study design where a focus on the Munich metropolitan area has been put. In contrast, the  nation-wide sample matches the overall population quite well with only minor percentage point differences. Consequently, we can conclude that from a spatial perspective the nation-wide sample can be considered representative. 

\begin{table}
\centering
\caption{Share of respondents in the postal regions of Germany in comparison to the overall population distribution in Germany. The location of the respective postal code area is shown in Figure \ref{fig:homelocations}.}\label{tab:1}
    \begin{tabular}{lllllllllll} \toprule
     Postal code region & 0 & 1 & 2 & 3 & 4 & 5 & 6 & 7 & 8 & 9 \\
     \midrule
    \multicolumn{11}{l}{Overall population distribution} \\
     \% of respondents & 7.9\ \% & 8.4\% & 10.5\% & 10.7\% & 12.3\% & 11.0\% & 9.1\% & 10.5\% & 9.3\% & 8.5\% \\
     \midrule
    \multicolumn{11}{l}{Pooled sample} \\
         \% of respondents & 3.7\ \% & 3.7\ \% & 4.8\ \% & 4.1\ \% & 6.0\ \% & 5.2\ \% & 5.2\ \% & 5.1\ \% & 55.7\ \% & 6.8\ \% \\
     \midrule
     \multicolumn{11}{l}{Nation-wide sample (survey only)} \\
         \% of respondents & 7.8\ \% & 7.4\ \% & 10.8\ \% & 9.1\ \% & 14.0\ \% & 10.7\ \% & 11.0\ \% & 10.2\ \% & 7.8\ \% & 11.1\ \% \\
     \midrule
    \multicolumn{11}{l}{Munich-oriented sample (survey only)} \\
         \% of respondents & 0.8\ \% & 1.9\ \% & 0.4\ \% & 0.8\ \% & 0.8\ \% & 2.3\ \% & 1.9\ \% & 3.5\ \% & 82.6\ \% & 5.0\ \% \\
    \midrule
    \multicolumn{11}{l}{Munich-oriented sample (app+survey)} \\
         \% of respondents & 0.8\ \% & 0.9\ \% & 0.7\ \% & 0.6\ \% & 0.4\ \% & 1.1\ \% & 0.9\ \% & 1.0\ \% & 90.1\ \% & 3.54\ \% \\
    \midrule
    \end{tabular}
\end{table}

\begin{figure}
    \centering
     \begin{subfigure}[b]{0.48\textwidth}
         \centering
         \includegraphics[width=\textwidth]{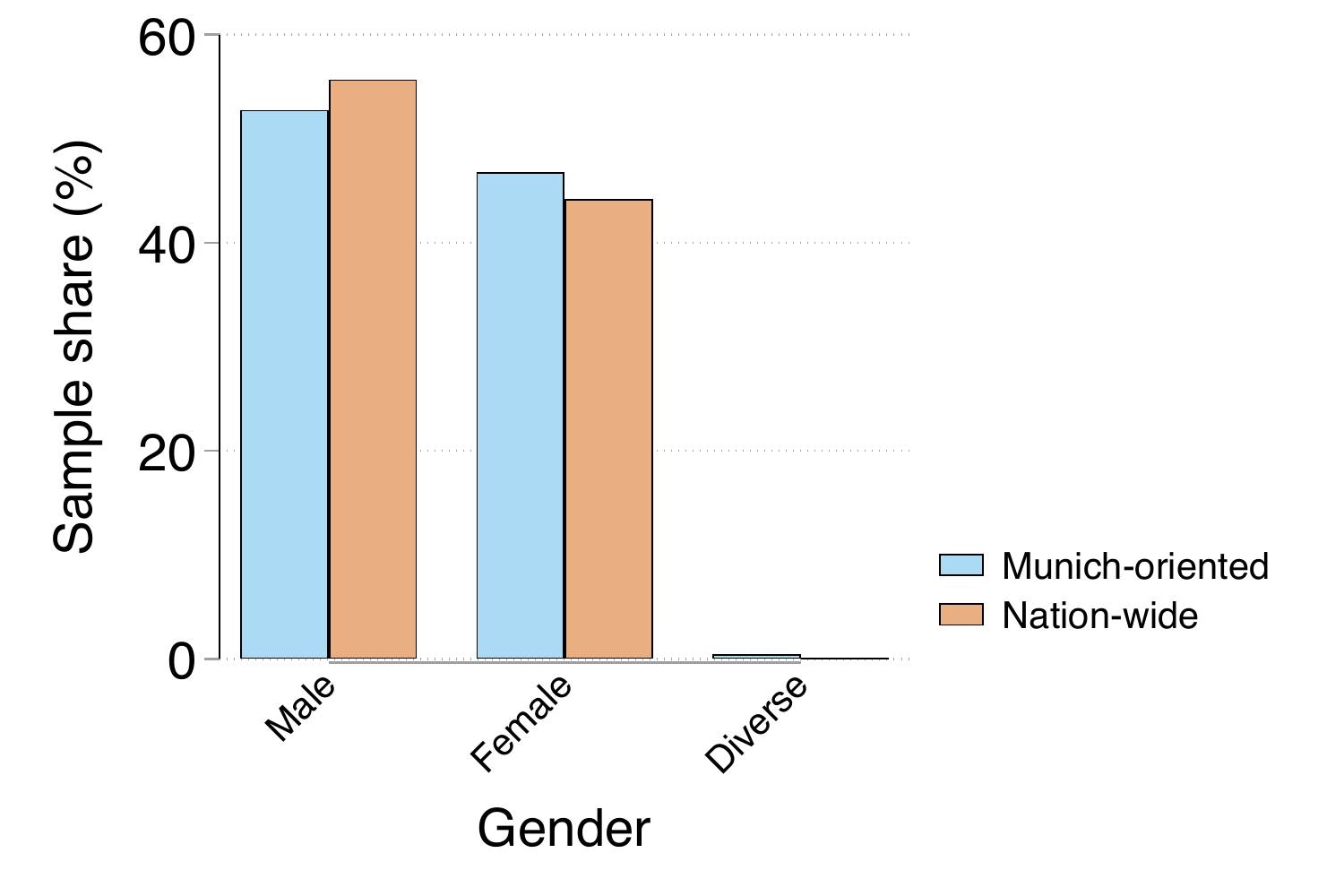}
         \label{fig:age}
     \end{subfigure}
    \hfill
     \begin{subfigure}[b]{0.48\textwidth}
         \centering
         \includegraphics[width=\textwidth]{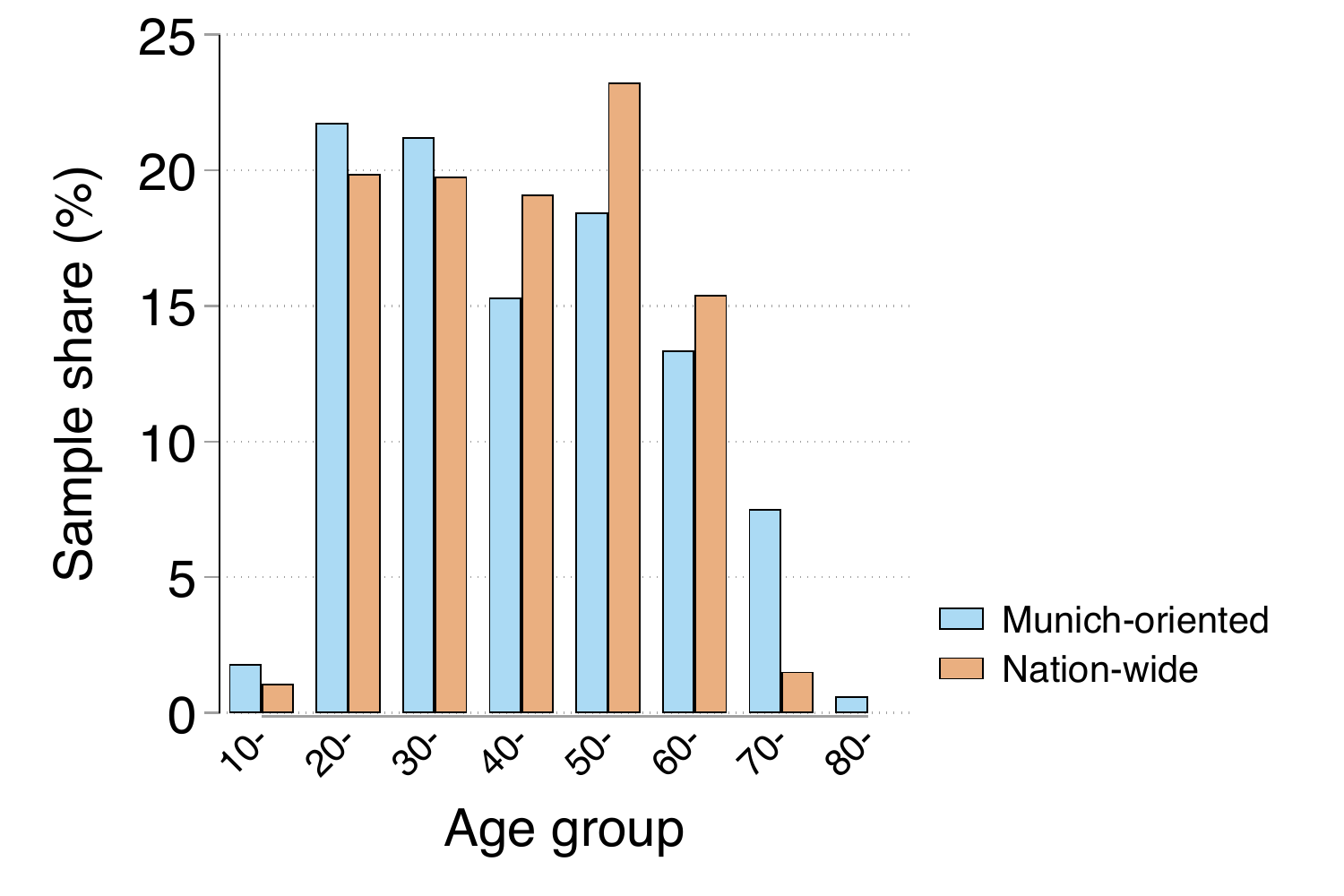}
         \label{fig:sex}
     \end{subfigure}

     \begin{subfigure}[b]{0.48\textwidth}
         \centering
         \includegraphics[width=\textwidth]{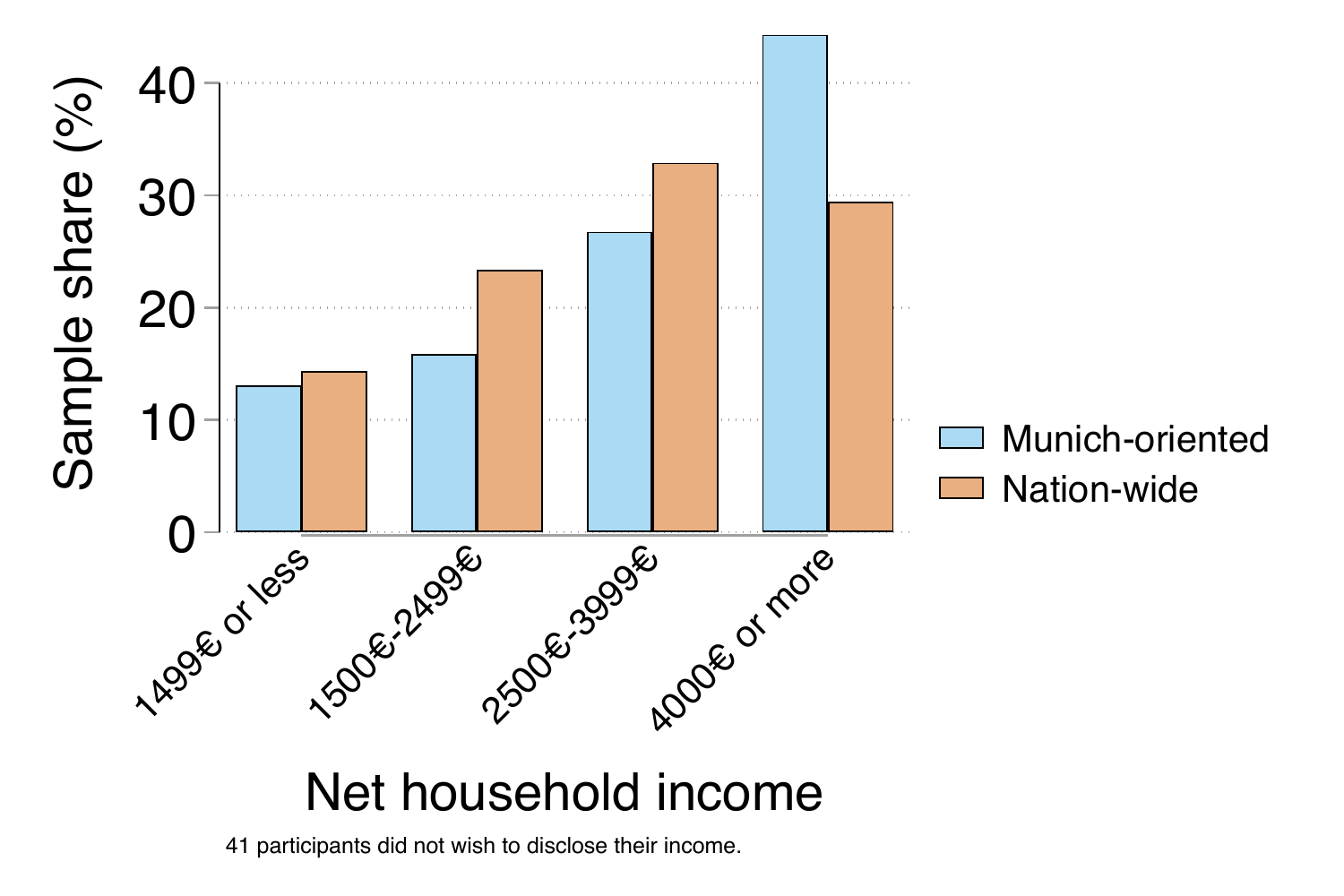}
         \label{fig:sample_income}
     \end{subfigure}
    \hfill
     \begin{subfigure}[b]{0.48\textwidth}
         \centering
         \includegraphics[width=\textwidth]{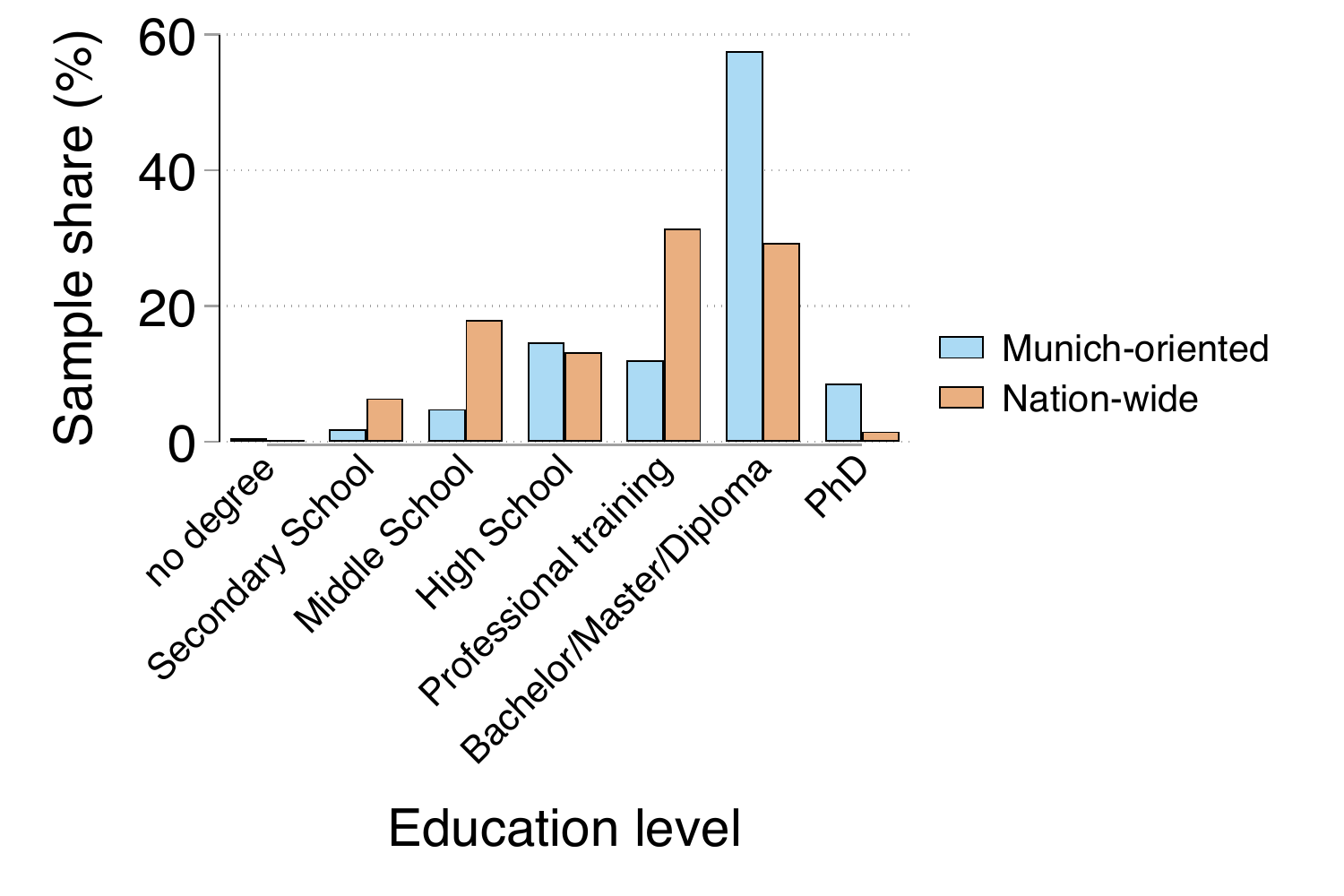}
         \label{fig:education}
     \end{subfigure}
    \caption{Socio-economic characteristics of the sample.}
    \label{fig:socioeco}
\end{figure}

The socio-economic characteristics of the sample are shown in Figure \ref{fig:socioeco}. There are slightly more male respondents in the nation-wide and the Munich oriented sample with a share of around 54\ \%. Thus, male respondents are slightly over-represented in our entire study as the overall population has roughly an equal share of males and females. The average age in the pooled sample is around 44.3 years, while the nation-wide and Munich-oriented sample having almost identical values. This value is very close to the average age of the overall German population (44.6 years), but note that our sample does not contain children and some of the elderly. Nevertheless, it becomes apparent that our Munich-oriented sample is over-sampling in the age category of 20-40, while under-sampling the baby-boomer generations around 50-60.

In 2020, the average net household income was around 3\ 600\ EUR and the median net household income was around 2\ 950\ EUR\footnote{Wirtschaftsrechnungen, Laufende Wirtschaftsrechungen. Einkommen, einnahmen und Ausgaben privater Haushalte. Fachserie 2015 Reihe 1, DESTATIS, Statistisches Bundesamt, 2021.}. Our sample as shown in Figure \ref{fig:socioeco} seems to match these values at least for the nation-wide sample, considering that the median of this sample is somewhere between 2\ 500 and 3000\ EUR. Nevertheless, the Munich-oriented sample is clearly biased towards the upper end, not only caused by the higher wages in this region of Germany, but also as this sample is substantially over-sampling people with an academic degree, i.e., around 20\ \% in the overall population \footnote{\url{https://www.destatis.de/DE/Themen/Gesellschaft-Umwelt/Bildung-Forschung-Kultur/Bildungsstand/_inhalt.html}} compared to almost 60\ \% in the Munich-oriented sample. However, note that the offical figures also includes children which is likely to increase the share of academics a little. In contrast, the nation-wide sample matches the overall figures better with around 30\ \% of the sample.



%% file: _sections/mobility.tex
\begin{figure}
    \centering
     \begin{subfigure}[b]{0.48\textwidth}
         \centering
         \includegraphics[width=\textwidth]{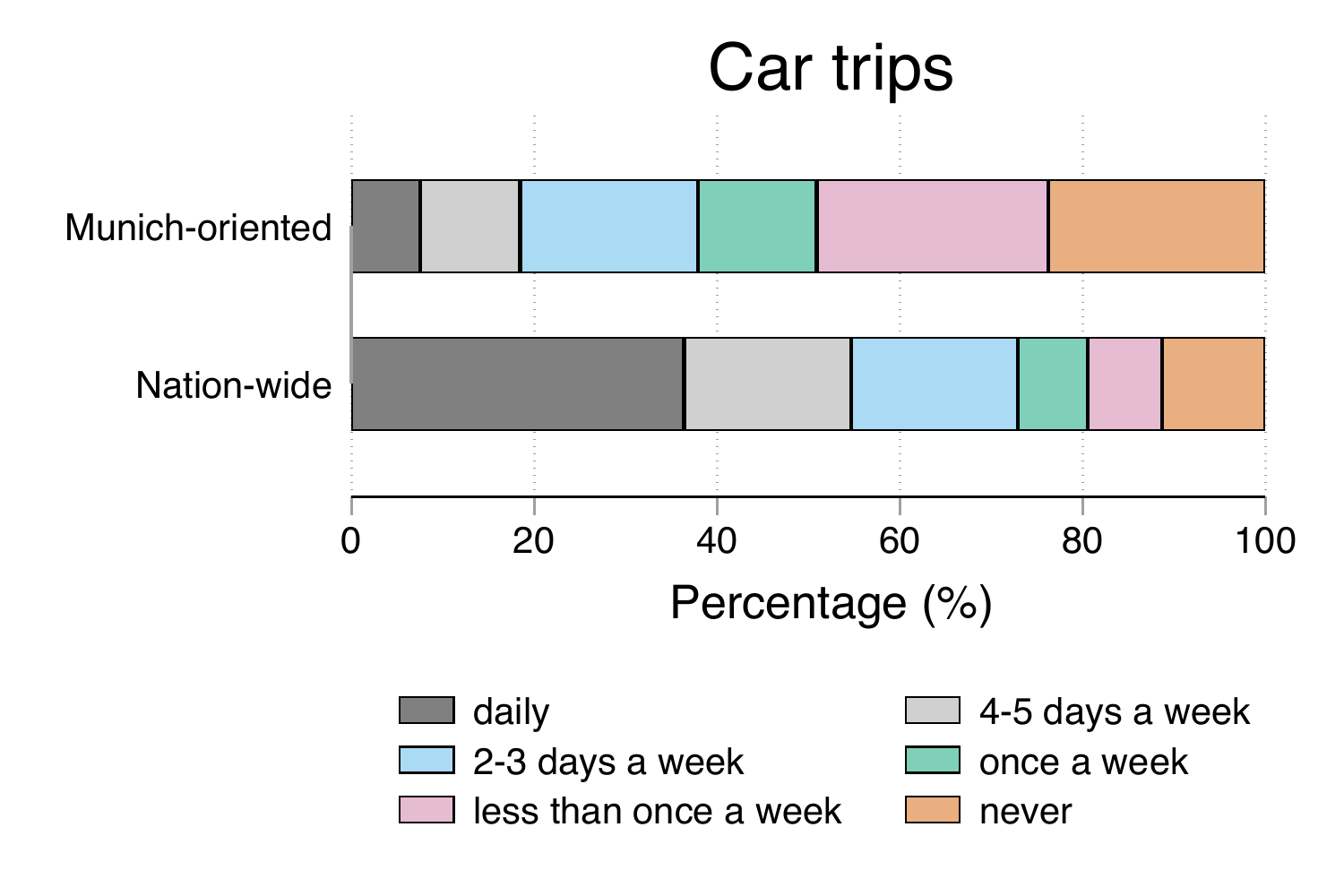}
         \label{fig:carfreq}
     \end{subfigure}
    \hfill
     \begin{subfigure}[b]{0.48\textwidth}
         \centering
         \includegraphics[width=\textwidth]{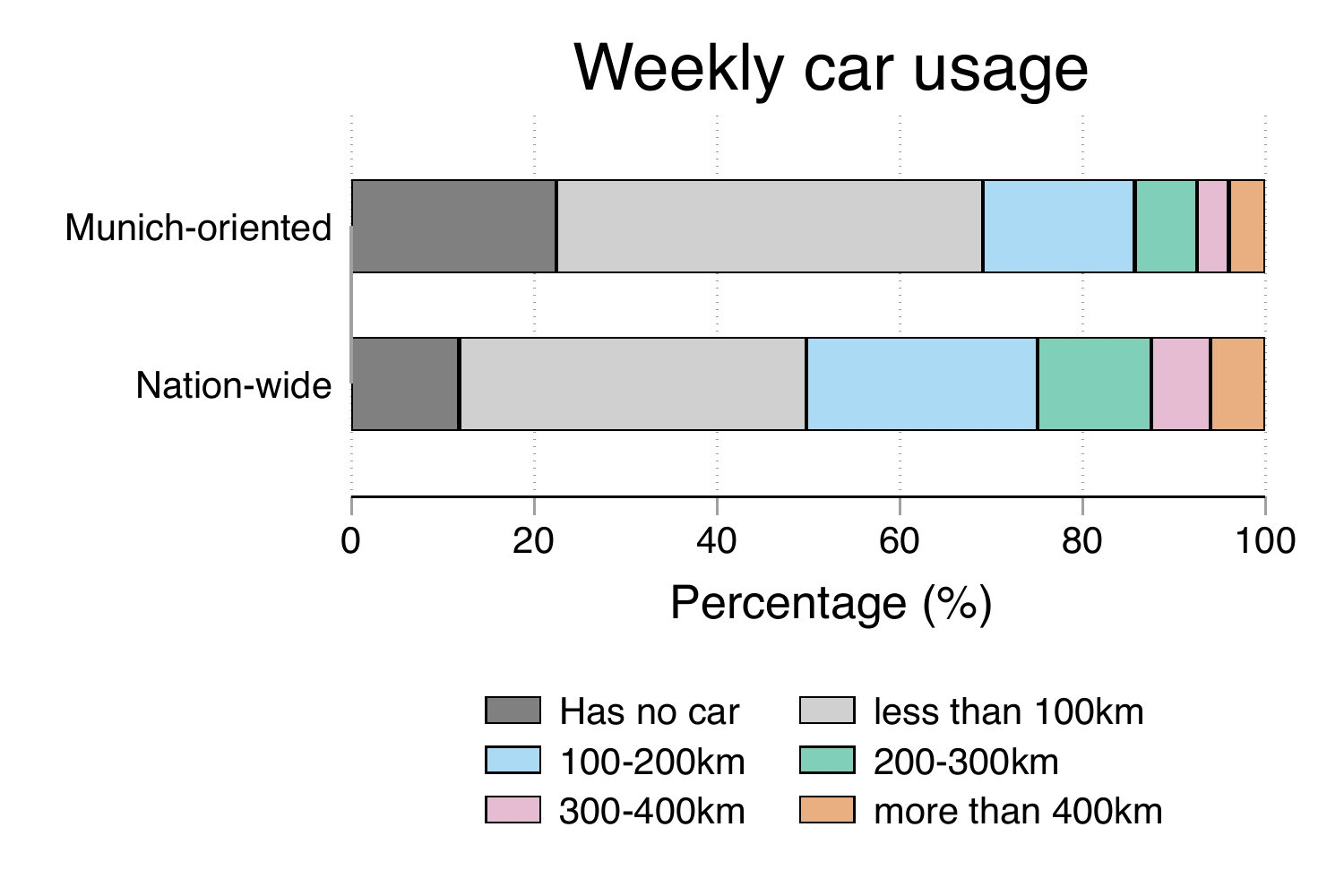}
         \label{fig:caruse}
     \end{subfigure}

     \begin{subfigure}[b]{0.48\textwidth}
         \centering
         \includegraphics[width=\textwidth]{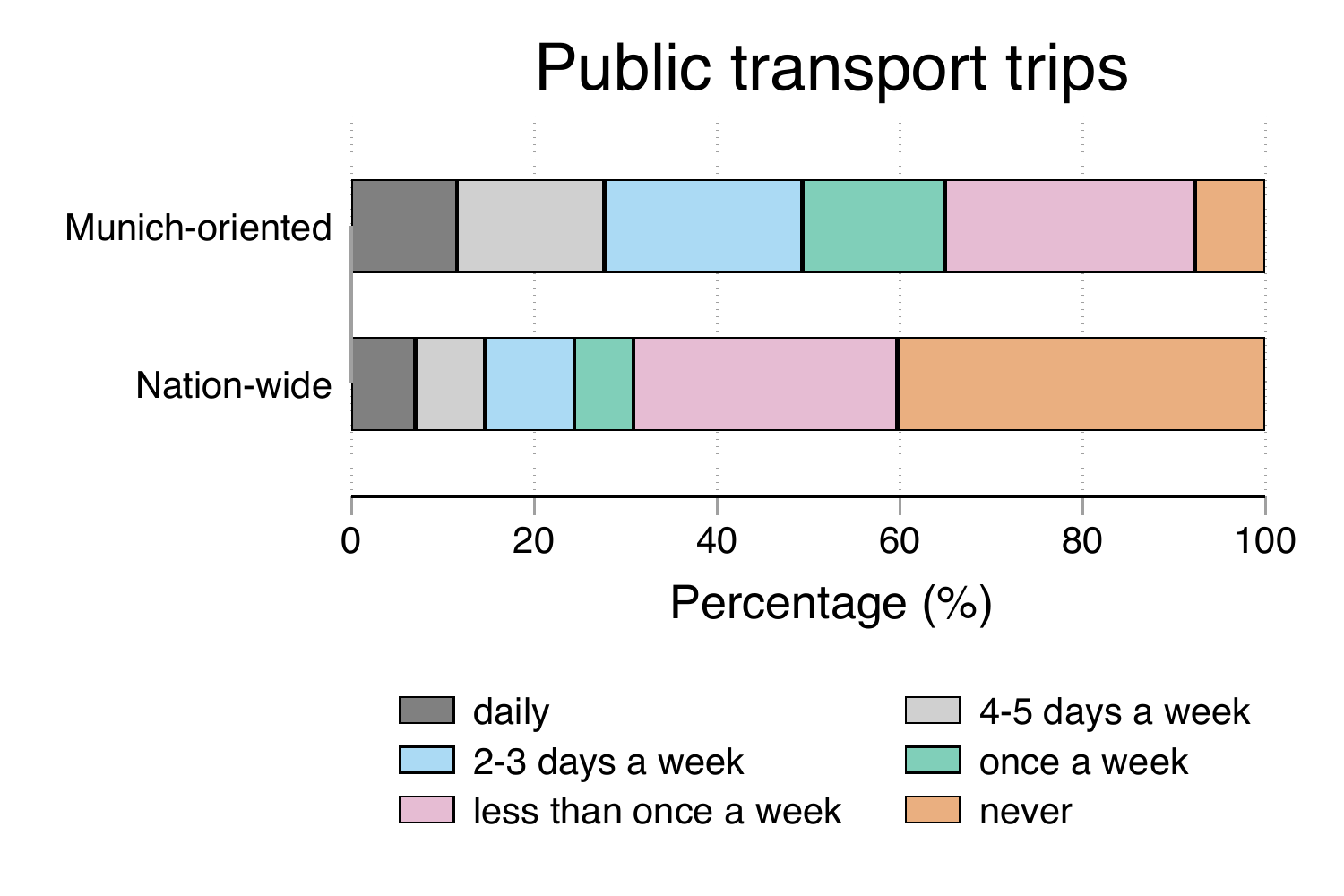}
         \label{fig:ptuse}
     \end{subfigure}
    \hfill
     \begin{subfigure}[b]{0.48\textwidth}
         \centering
         \includegraphics[width=\textwidth]{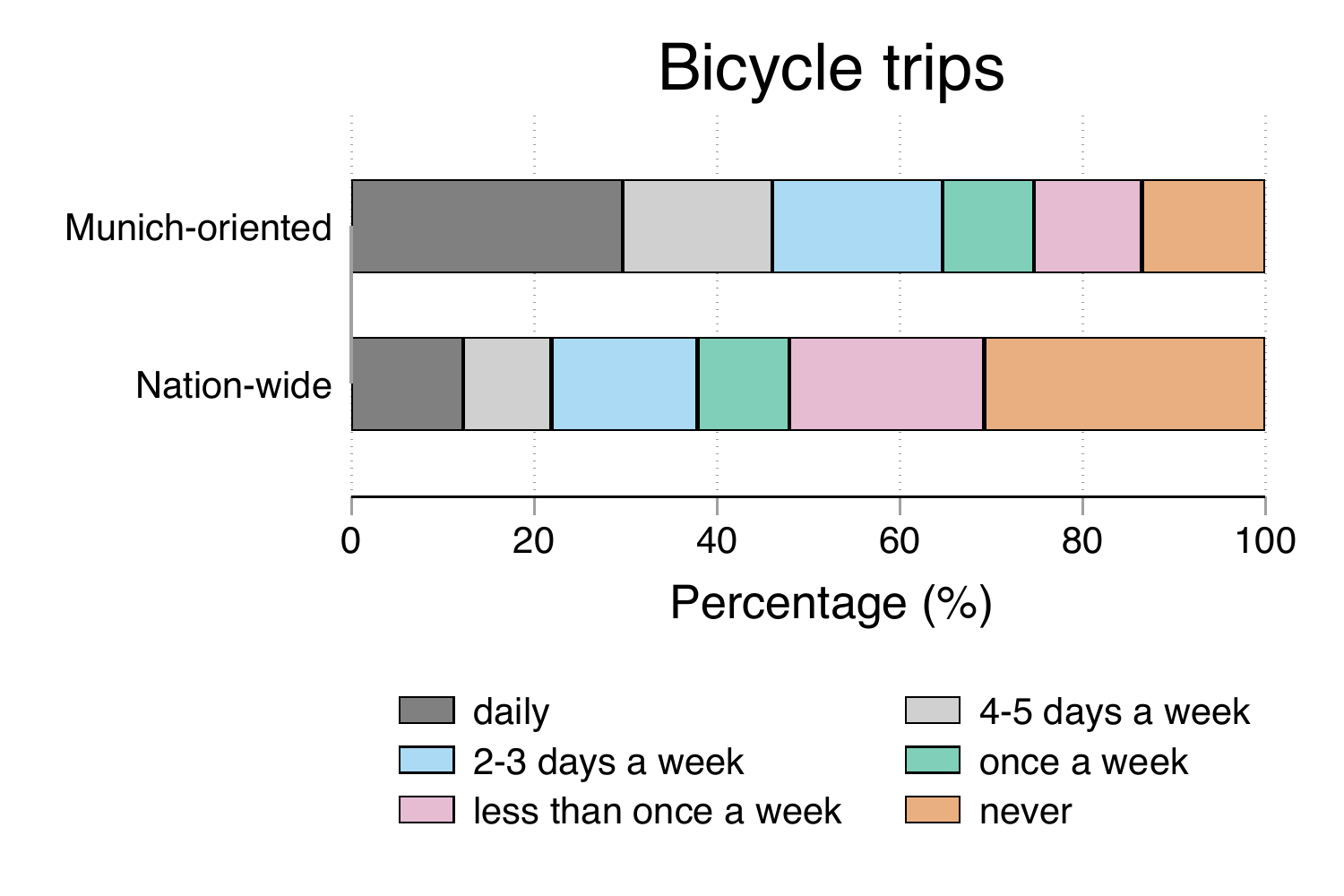}
         \label{fig:velofreq}
     \end{subfigure}
    \caption{Reported travel behavior in our study.}
    \label{fig:mob_results}
\end{figure}

%
%
%
%
%
%
We compare the stated travel behavior of the Munich-oriented and nation-wide sample of our study with the official figures for Germany from the official travel survey (MiD)\cite{dlr125879}. Fortunately, the same travel survey provides a detailed report on travel behavior in Munich and its suburban districts within the perimeter of the MVV, Munich's transit authority \cite{mid_munich2019}. Selected results from the first-wave survey are shown in Figure \ref{fig:mob_results} and are discussed in the following.

The MiD reports a car ownership rate in the MVV area of 31\ \% which is almost similar to the number we have in our Munich-oriented sample. In the nation-wide sample, we have 13.4\ \% individuals without a car, while the MiD reports 22\ \%. The share in the Munich-oriented sample that uses a car almost daily (on 4 to 7 days a week) is 18.5\ \%, which is substantially smaller than the 36\ \% reported by MiD for the MVV area and the reported 22\ \% for the Munich city area. The almost daily car usage in our nation-wide sample with around 54.7\ \% of the sample is slightly higher than the 50\ \% reported by MiD for Germany. The share of individuals who never use a car is 23.7\ \% in the Munich-oriented sample, which is larger than the MiD figure with 16\ \% for the MVV area and 22\ \% for the city of Munich. The share of individuals in our nation-wide sample that never uses the car is 11.3\ \%, which fits perfectly the figure reported in the MiD.

In our Munich-oriented sample, 60\ \% did not have a public transport subscription before the introduction of the 9\ EUR-ticket, e.g., a monthly pass. This is slightly less than the 65\ \% reported for the MVV area in the MiD. For the nation-wide sample, around 75\ \% stated not to have any public transport subscription that is seven percentage points less than reported in the MiD study. The share of people using public transport almost daily in our Munich-oriented sample is 27.7\ \%, nearly identical with the MiD figure of 26\ \% for the MVV area. In comparison, the share for the city of Munich is 34\ \%. For our nation-wide sample, we find a reported share of 14.7\ \% that is almost two percentage points more than in the MiD's report. The share of people in our Munich-oriented sample that never uses public transport is  7.7\ \%; comparing to the MiD results, this value is very close to reported values for the city of Munich, but smaller than for the MVV area at 16\ \%. We find that the share of individuals in our nation-wide sample that never uses public transport is 40.3\ \%, which matches perfectly the values reported by in the MiD study.

Regarding bicycle usage, we find that the share for almost daily usage is 46.1\ \% in our Munich-oriented sample that is substantially larger than the MiD values of 28\ \% for the city of Munich and 26\ \% for the MVV area. The nation-wide sample reported a share of 21,9\ \%, which is slightly larger than the 18\ \% reported in the MiD study for Germany. In our Munich-oriented sample, the share of individuals who never use a bicycle is 13.4\ \%; this is much smaller than the numbers reported in the MiD study for the city of Munich (26\ \%) and the MVV area (25\ \%). Also for our nation-wide sample we find a similar pattern, where the reported share is 30.7\ \%, six percentage points smaller that reported in the MiD study. 

We see that our nation-wide sample matches the MiD figures quite well, despite comparing pre- and post-COVID-19 figures. Nevertheless, one can observe differences between our sample and the MVV area reported in the MiD. The differences can be explained by having not the same spatial extent and that our sample is not yet weighted. However, it can already be expected that we over-sampled the Munich urbanized area and in particular cyclists, albeit it cannot be ruled out that this differences also result from post-COVID travel behavior and the summer season the survey took part.





%% file: _sections/energy.tex
The first wave of the survey also covers the aspects of energy consumption and the household expenditures to account for the price development in the previous months. In the following, selected results are shown.

\begin{figure}
     \centering
     \begin{subfigure}[b]{0.48\textwidth}
         \centering
         \includegraphics[width=\textwidth]{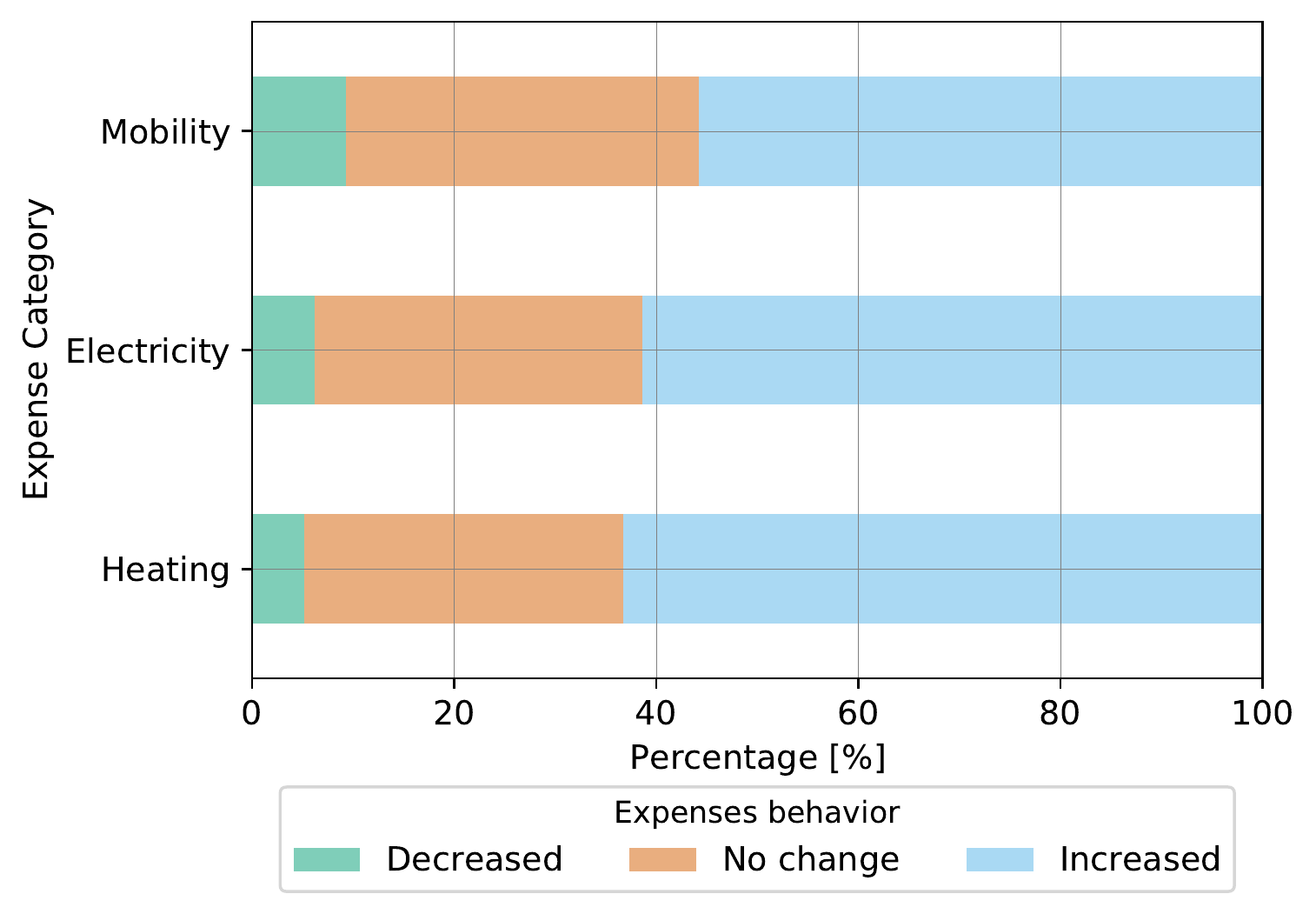}
         \caption{Behavior of expenses}
         \label{fig:expense_change}
     \end{subfigure}
     \hfill
     \begin{subfigure}[b]{0.48\textwidth}
         \centering
         \includegraphics[width=\textwidth]{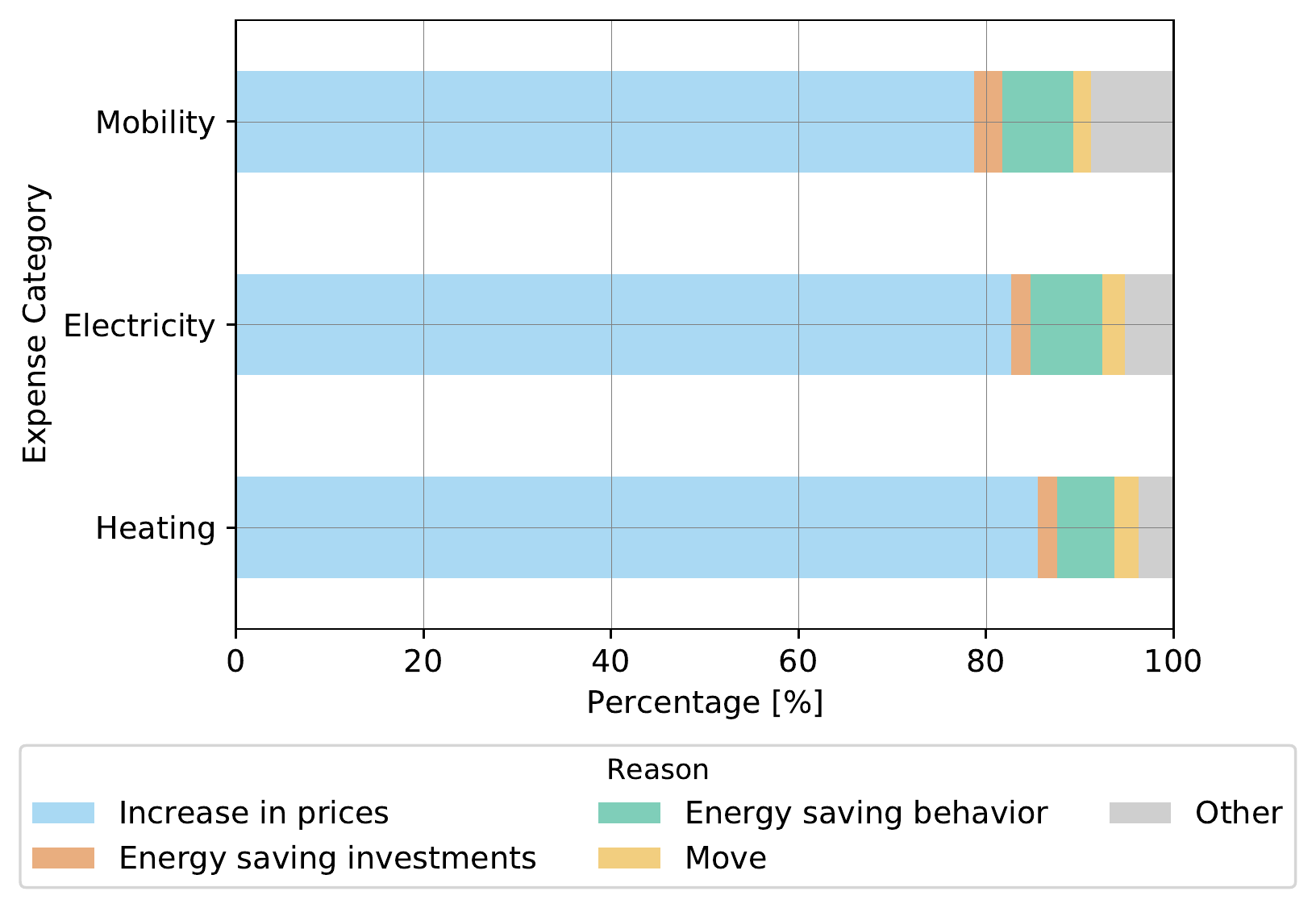}
         \caption{Reasons of cost increase or decrease.}
         \label{fig:expense_change_why}
     \end{subfigure}
     
     \begin{subfigure}[b]{0.48\textwidth}
         \centering
         \includegraphics[width=\textwidth]{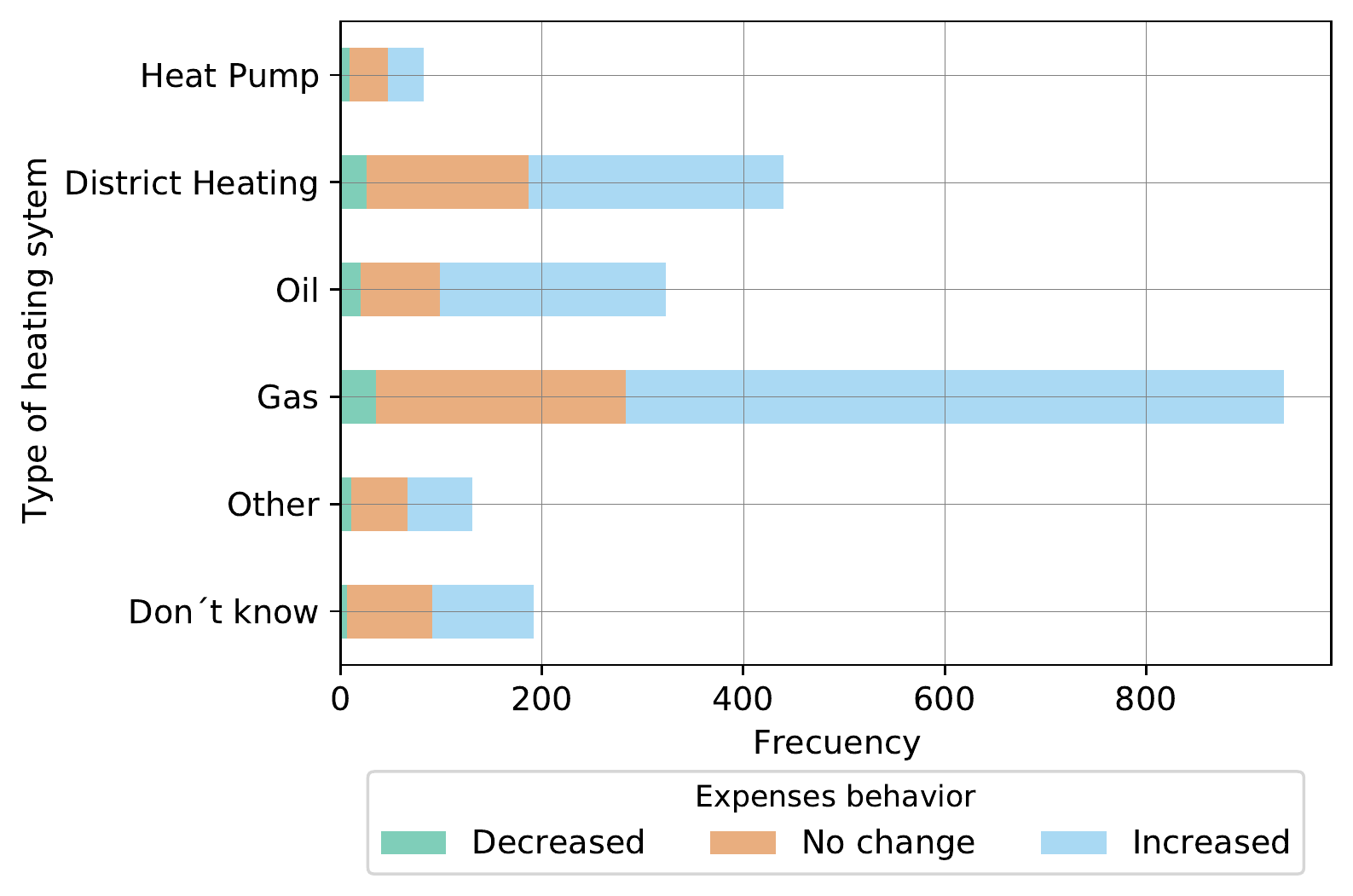}
         \caption{Household heating system type}
         \label{fig:heat_type}
     \end{subfigure}
     \hfill
     \begin{subfigure}[b]{0.48\textwidth}
         \centering
         \includegraphics[width=\textwidth]{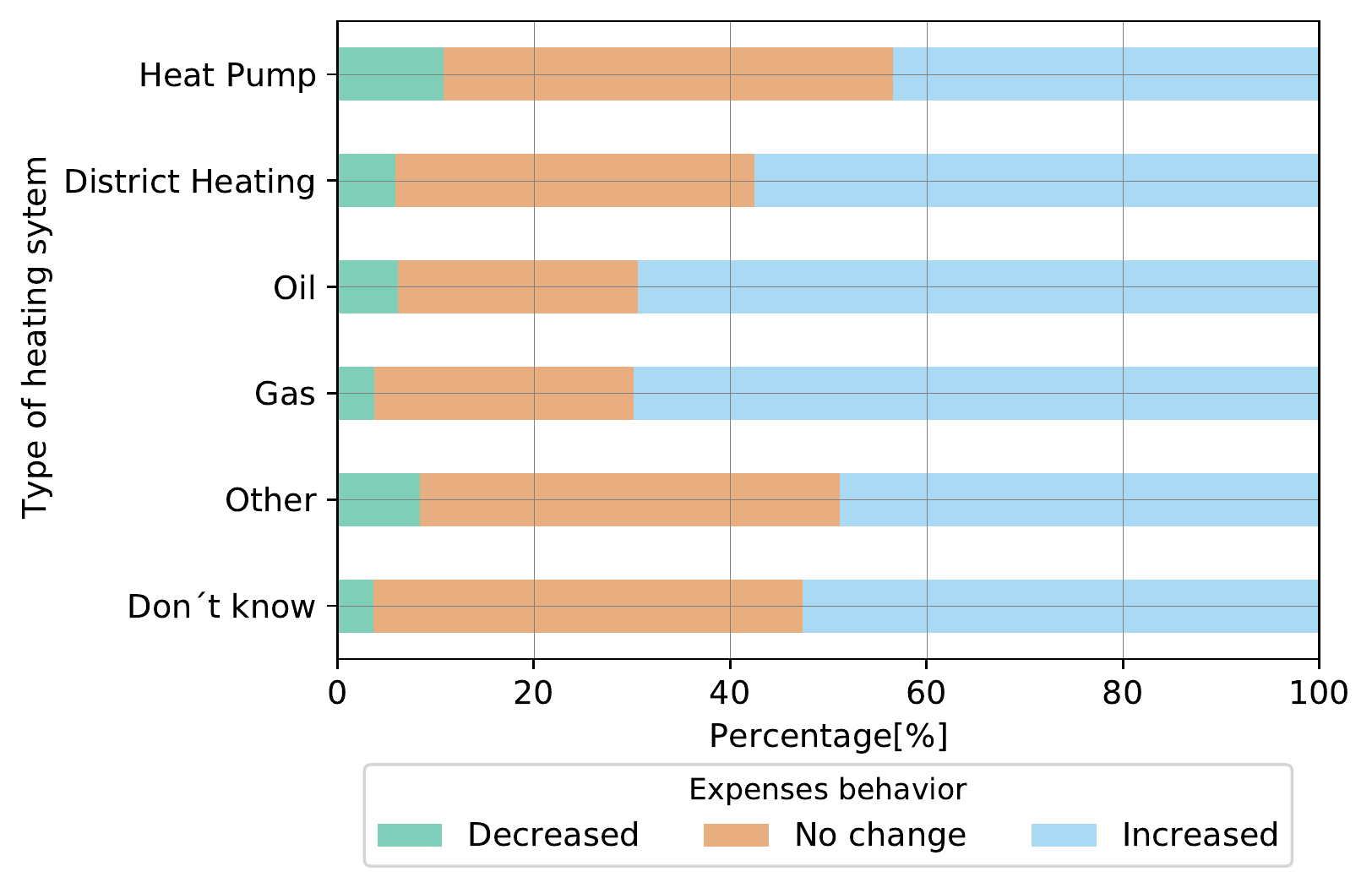}
         \caption{Distribution cost behavior for the heating system types}
         \label{fig:heat_percent}
     \end{subfigure}
     \caption{Energy expenses of the pooled sample.}
     \label{fig:energy_basic}
\end{figure}

The respondents were asked whether their costs for heating, electricity and mobility had changed in the past few months. Figure \ref{fig:expense_change} shows that across all three dimensions, respondents experienced an increase, specifically 63\% of the respondents reported an increase in their expenses for heating, 61\% for electricity, and 56\% in mobility. Less than 10\% of them said that their costs have been reduced. The remaining share of respondents reported no significant change in their costs. 

If a respondent indicated a change in costs in either of the three categories of mobility, electricity and heating, he or she was asked about the causes for each reported change. As seen in Figure \ref{fig:expense_change_why}, the majority of respondents ($>75$\ \%), who reported a change, said it was caused by the increase in prices in all categories. This was particularly evident in the costs for heating where 86\ \% of the affected respondents said it was due to price increases; with natural gas prices seeing probably the most substantial price increases in recent months, it is not surprising that those respondents who use this heating source in our pooled sample (see Figure \ref{fig:heat_type}), are predominantly affected by price increases compared to other heating systems as seen in Figure \ref{fig:heat_percent}. When asked about the burden that results from the cost increases, only 20\ \% of households reported that these cost increases are not a burden to them, while around 30\ \% stated that it is a great burden, while the remaining share stated that it is a burden to them. When asked about how households change their economic activities in response to the cost increases, around 30\ \% reported that they are not affected, 40\ \% stated that they will spend less on leisure activities, 30\ \% indicated to postpone long-term investments and 36\ \% of respondents reported to save on everyday errands. Across the three latter categories we find consistent income effects where more affluent households are less likely to change their economic activities. 

\begin{figure}
    \centering
     \begin{subfigure}[b]{0.48\textwidth}
         \centering
         \includegraphics[width=\textwidth]{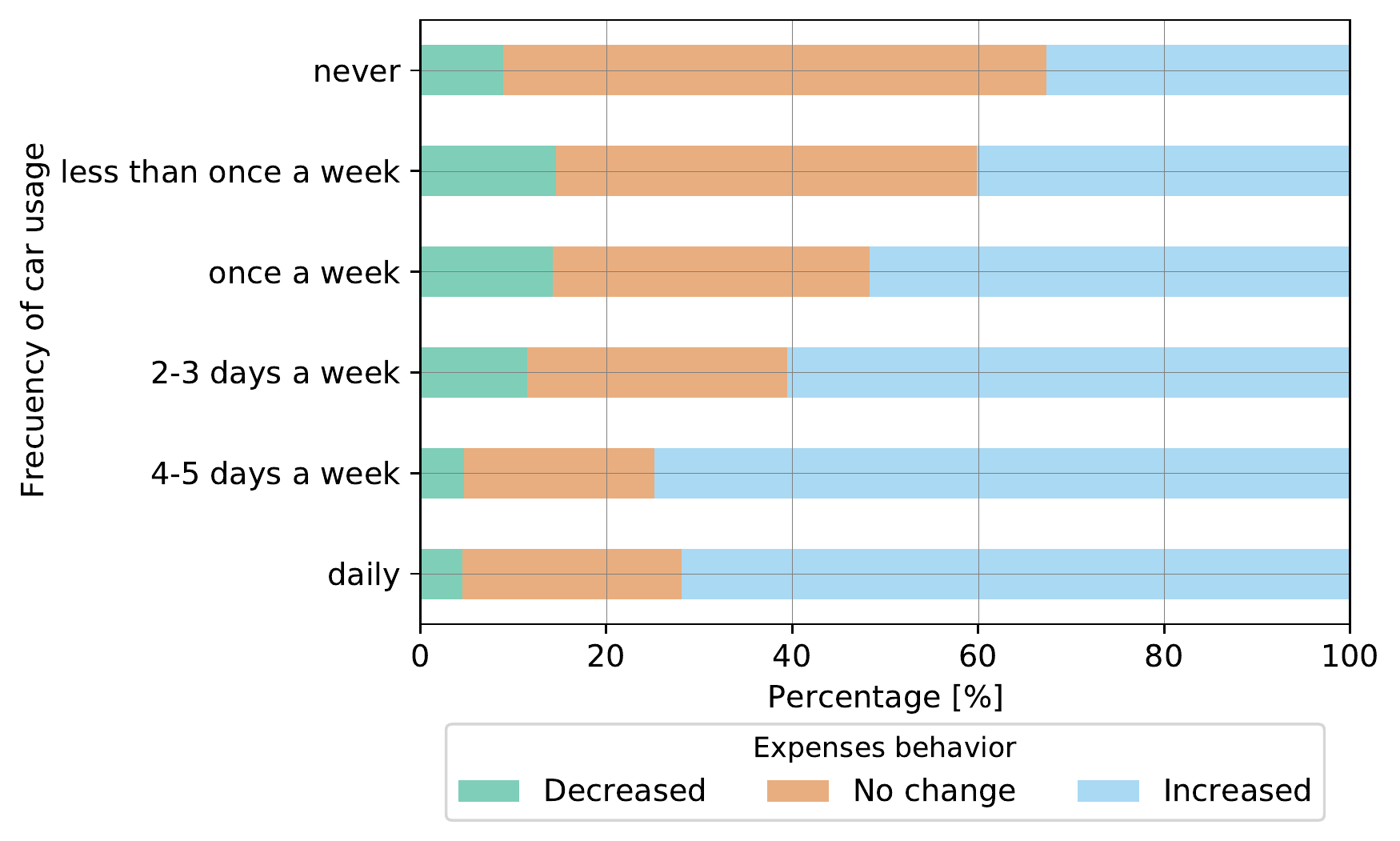}
         \caption{Car usage.}
         \label{fig:caruse_exp}
     \end{subfigure}
    \hfill
     \begin{subfigure}[b]{0.48\textwidth}
         \centering
         \includegraphics[width=\textwidth]{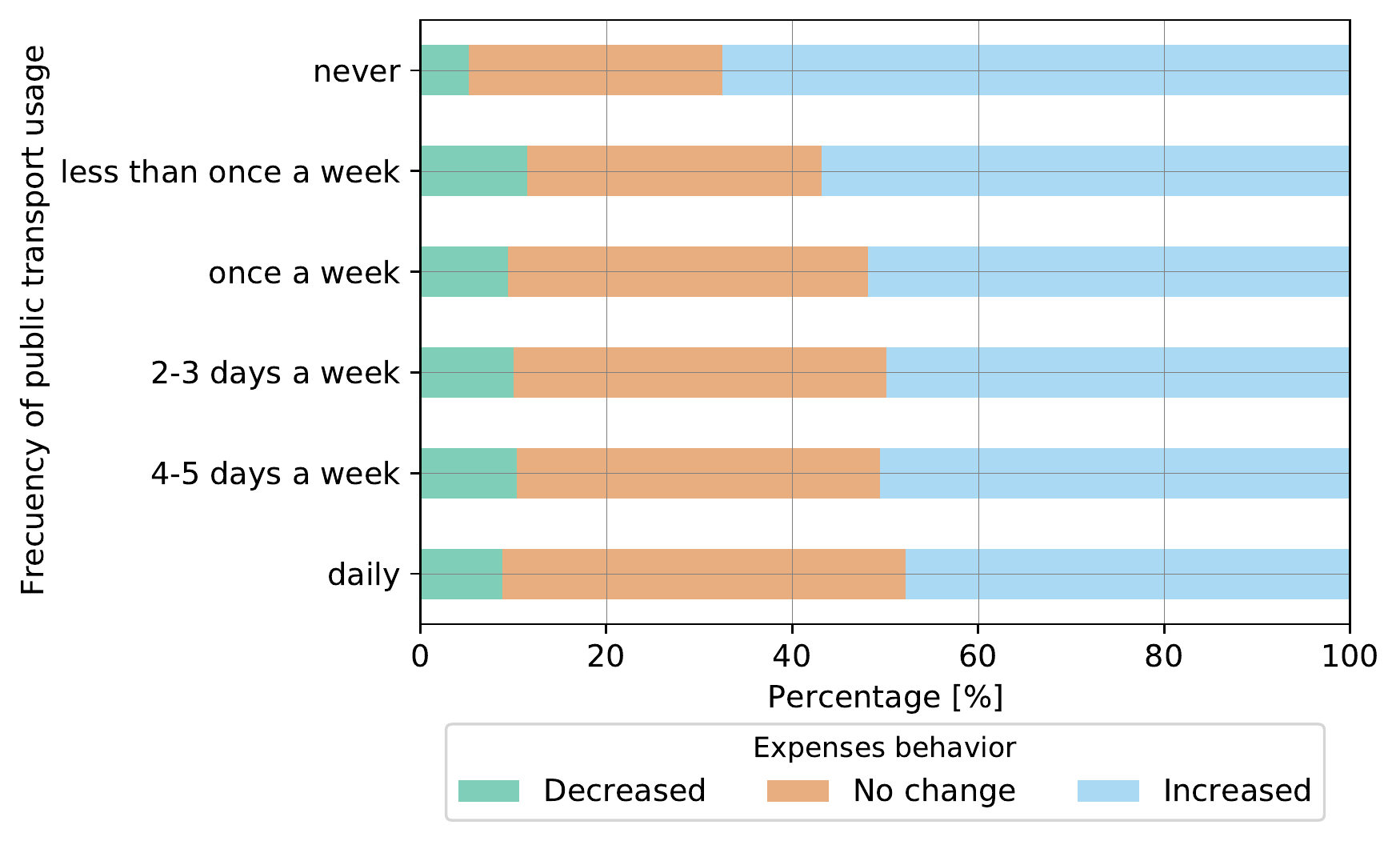}
         \caption{Public transport usage.}
         \label{fig:ptuse_exp}
     \end{subfigure}

     \begin{subfigure}[b]{0.48\textwidth}
         \centering
         \includegraphics[width=\textwidth]{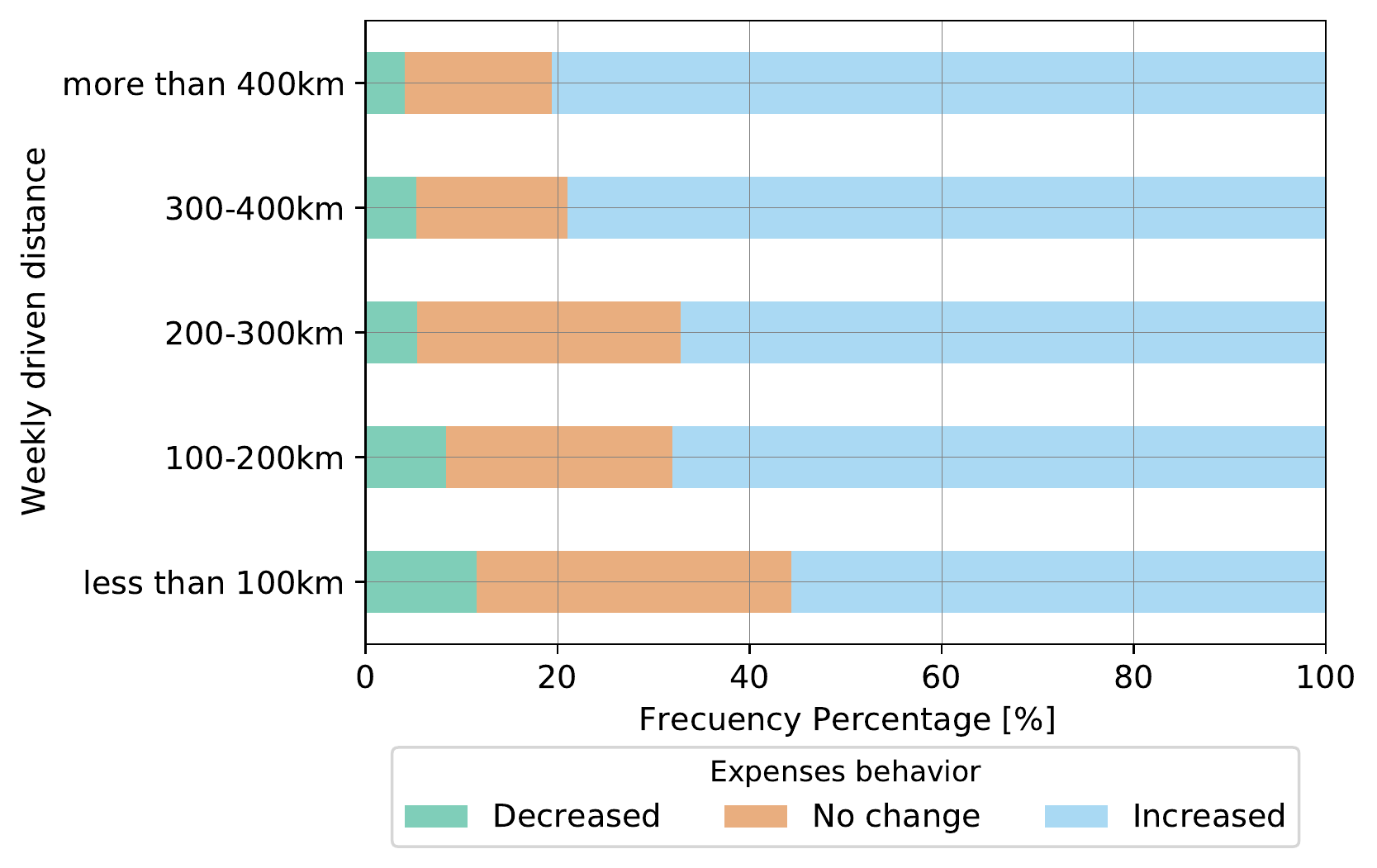}
         \caption{Weekly driven distance.}
         \label{fig:km_exp}
     \end{subfigure}
      \caption{Breakdown of cost changes by travel behavior variables.}
      \label{fig:cost_travelbehavior}
\end{figure}

For the mobility expense category, e.g., for fuels, car maintenance or public transport subscription, the experienced cost changes vary with travel behavior as seen in Figure \ref{fig:cost_travelbehavior}. As already seen with heating, the main increase in costs in mobility is due to the use of fossil fuels for cars: Figure \ref{fig:caruse_exp} shows the relationship between the frequency of car usage and the experienced of cost change; as it is evident, the more frequent a car is used, the larger the share of reported cost increases.  This is also closely related with the findings from Figure \ref{fig:km_exp}, where we analyze the experienced cost changes by the weekly traveled distances by car. Here, it is interesting to see that ~27\%  respondents who never use a car also reported cost increases, while the majority in this category did not report substantial changes to their mobility costs. In Figure \ref{fig:ptuse_exp}, we can see the opposite behavior for public transport usage compared to car usage: the more often a respondent uses public transport the less likely it is that they experienced mobility cost increases.

%% file: _sections/Ticket.tex
One of the aims of this study is to analyze how people perceive the political intervention of the 9 EUR ticket, both from a perspective of a cost reduction measure and as of a mean to promote more sustainable mobility as well as to understand people's intention to acquire the 9 EUR ticket. In the first wave of our study, we therefore assess the perceptions towards the ticket and the buying intention \emph {before} the introduction of the 9 EUR ticket. As in the previous sections, we separate the analysis by the sample type: the Munich-oriented and the nation-wide sample. Again, the rationale is to see whether there is a difference between both samples as we assume that individuals who register themselves for such a study may have biased attitudes compared to a representative sample.

Support and the intention to buy the 9 EUR ticket likely depends on the attitudes towards this political intervention. If a large portion of our sample expects negative (e.g., financial ruin of public transport) rather than positive consequences (e.g., relief for households, support for public transport) of the ticket, we can hardly expect them to support or buy it.  


\emph{Attitudes towards the ticket}

\begin{figure}
     \centering
     \begin{subfigure}[b]{0.48\textwidth}
         \centering
         \includegraphics[width=\textwidth]{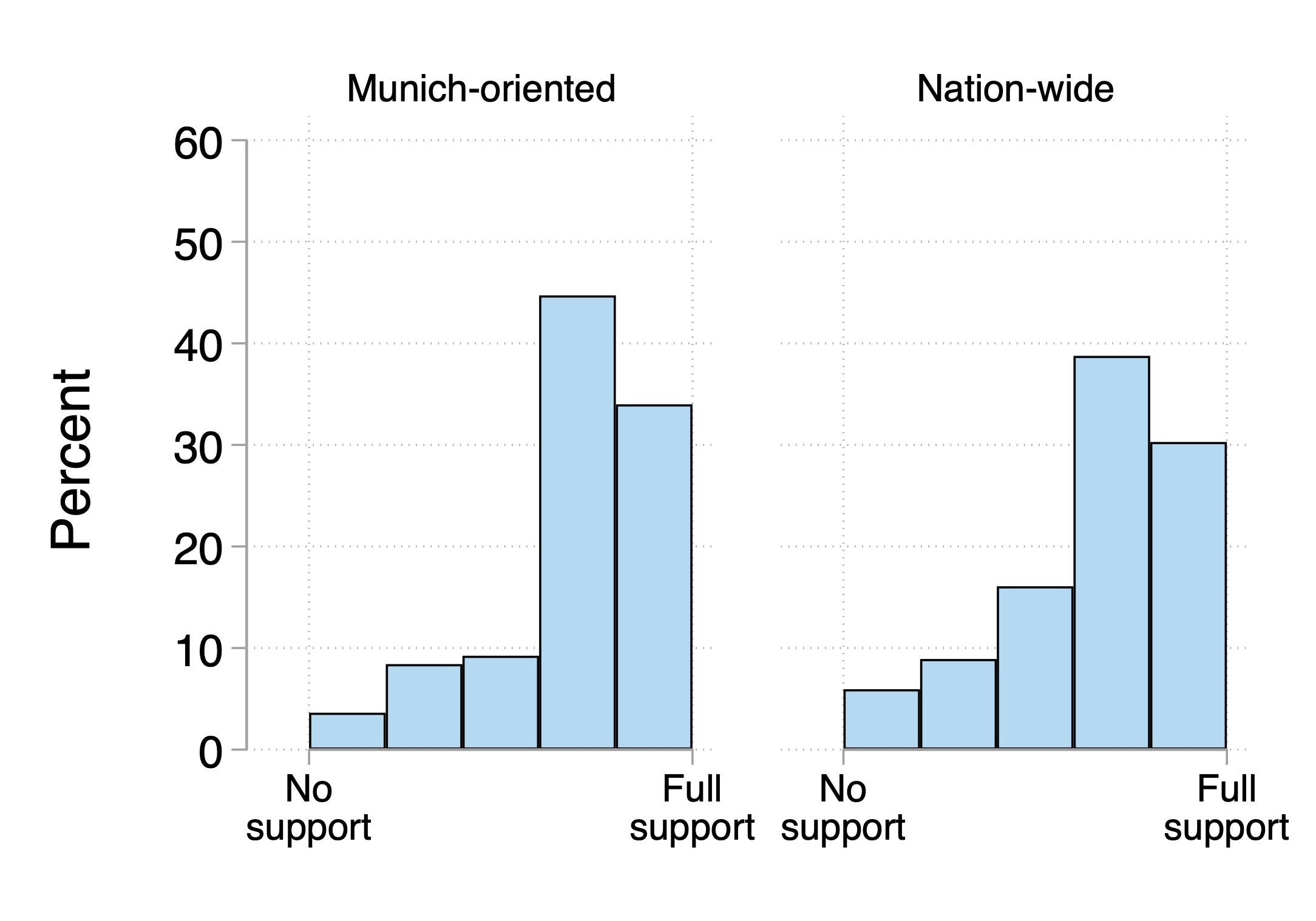}
         \caption{Ticket is a financial relief for households.}
         \label{fig:relief_hh}
     \end{subfigure}
     \hfill
     \begin{subfigure}[b]{0.48\textwidth}
         \centering
         \includegraphics[width=\textwidth]{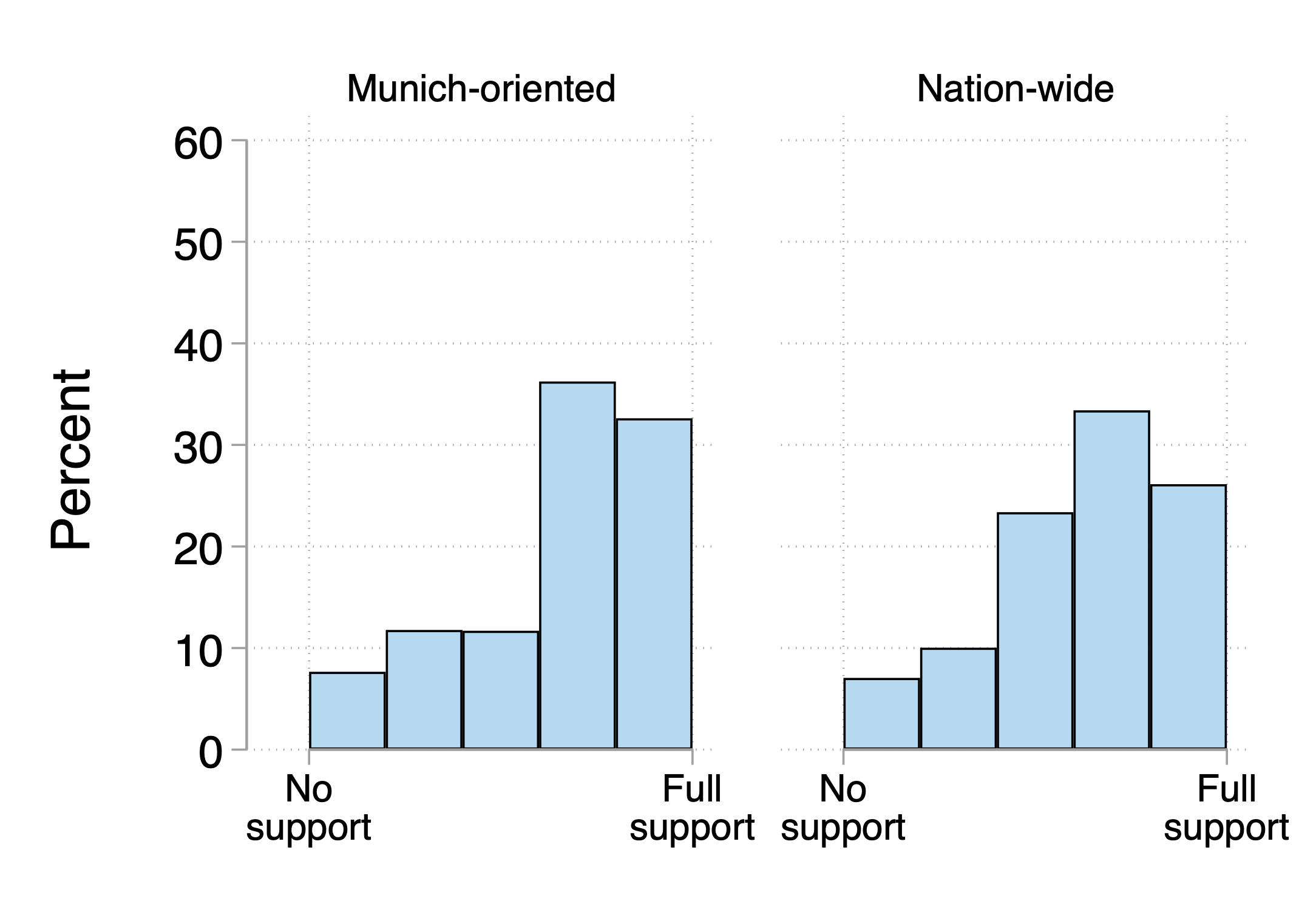}
         \caption{Ticket is important to support public transport.}
         \label{fig:support_pt}
     \end{subfigure}

     \begin{subfigure}[b]{0.48\textwidth}
         \centering
         \includegraphics[width=\textwidth]{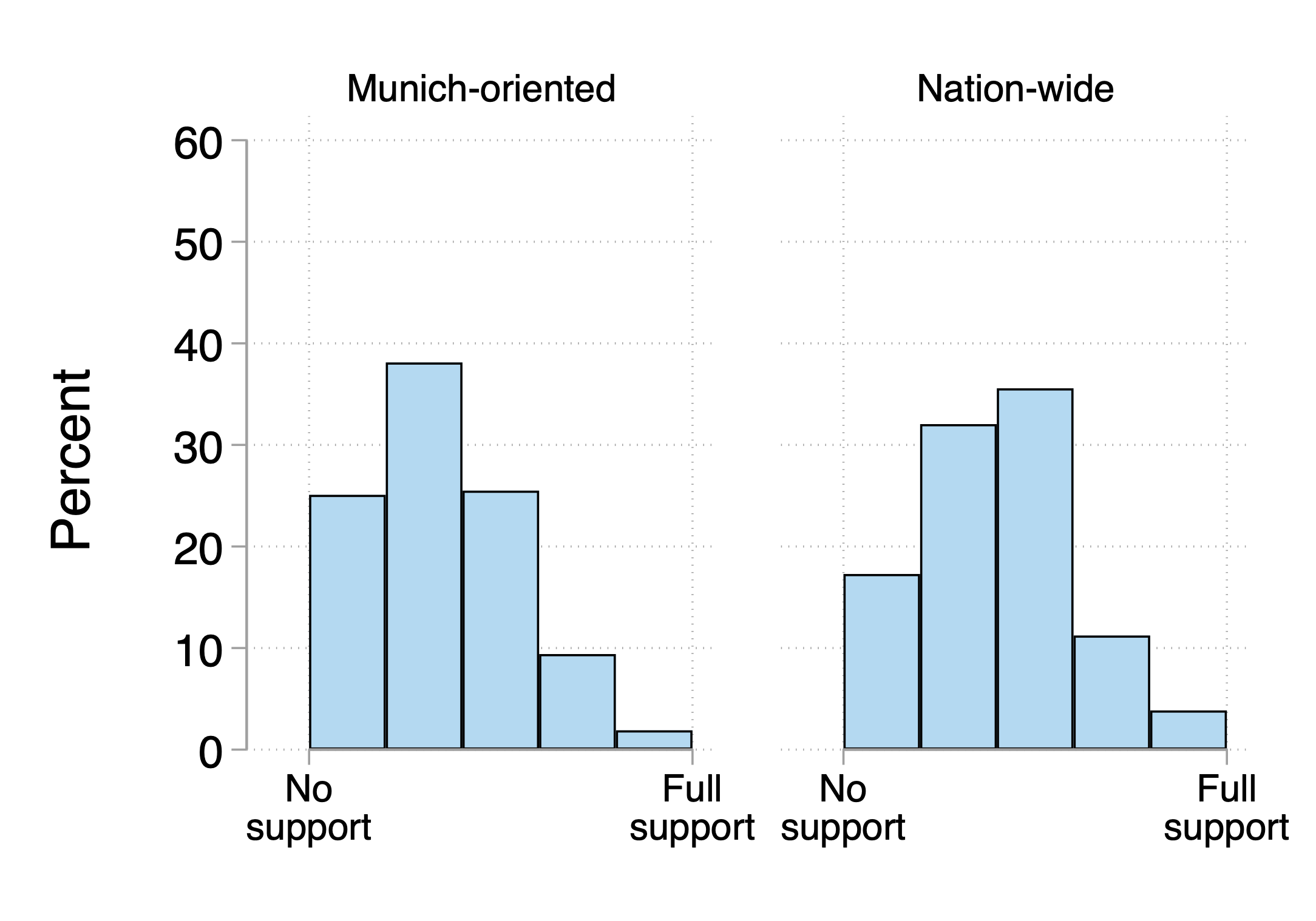}
         \caption{Ticket will ruin public transport financially.}
         \label{fig:ruin}
     \end{subfigure}
     \caption{Public attitudes towards the 9 EUR ticket.}
     \label{fig:ticket_attitudes}
\end{figure}

Looking at Figure \ref{fig:ticket_attitudes}, we find that respondents in both samples tend to have rather positive attitudes towards the ticket. For instance, we find that the majority of respondents in both samples at least agree that the ticket is a financial relief for households as seen in Figure \ref{fig:relief_hh}, and that it is important to support public transport as shown in Figure \ref{fig:support_pt}. In addition, a large share of respondents does not believe that the 9 EUR ticket will ruin public transport financially as seen in Figure \ref{fig:ruin}.

\emph{General support and intention to buy the ticket} 

\begin{figure}
     \centering
     \begin{subfigure}[b]{0.48\textwidth}
         \centering
         \includegraphics[width=\textwidth]{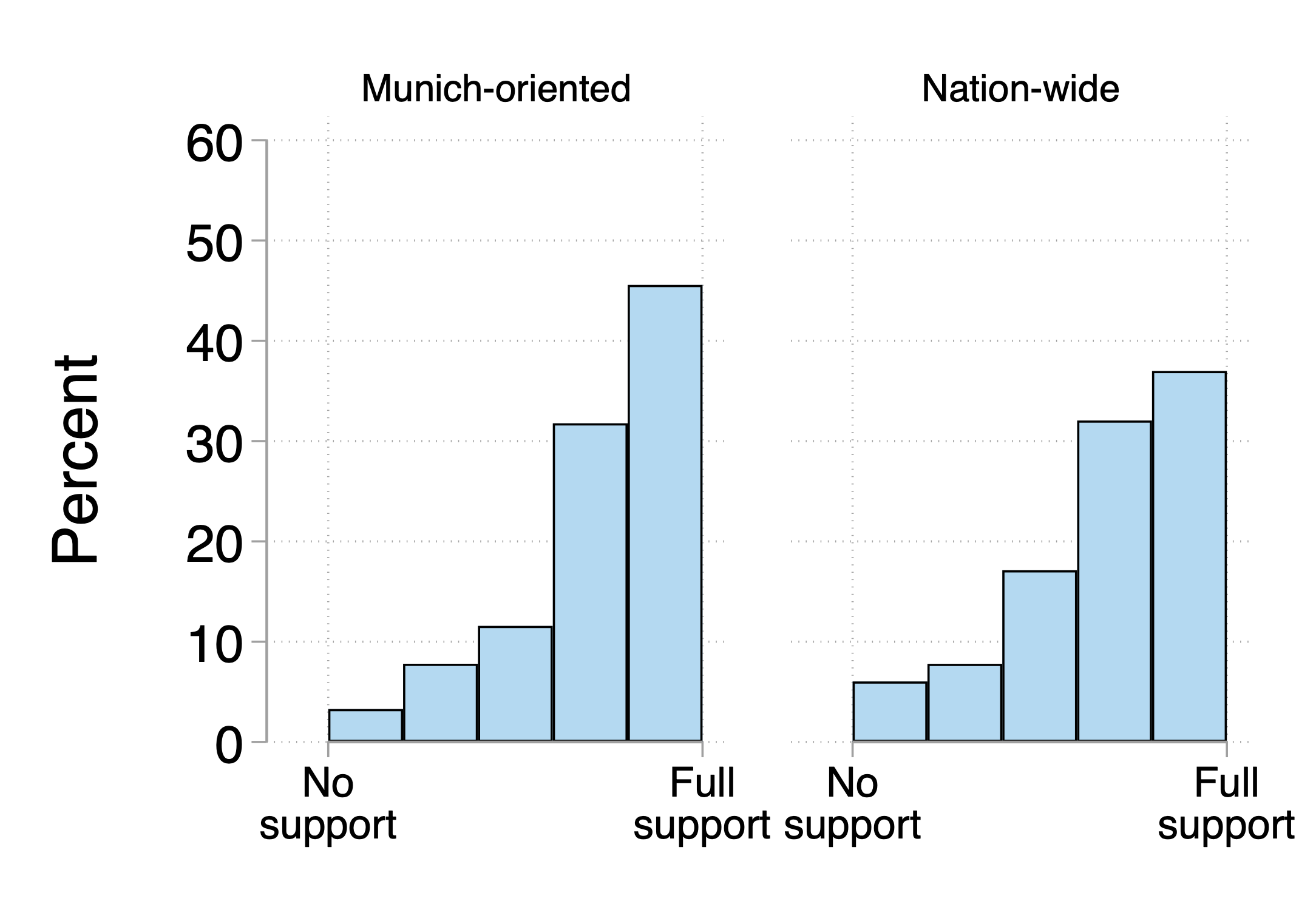}
         \caption{Support for ticket.}
         \label{fig:support}
     \end{subfigure}
     \hfill
     \begin{subfigure}[b]{0.48\textwidth}
         \centering
         \includegraphics[width=\textwidth]{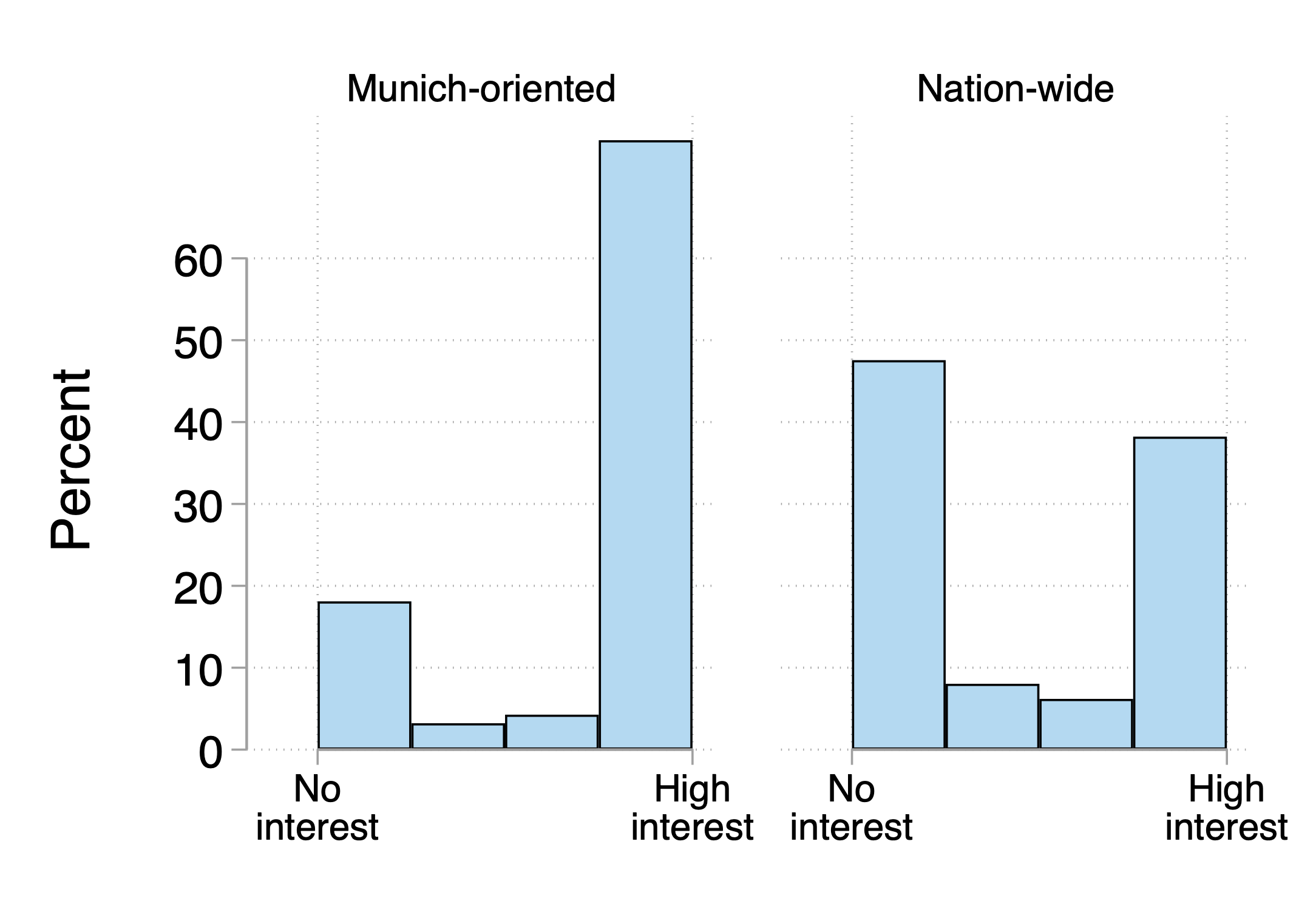}
         \caption{Overall interest in ticket.}
         \label{fig:interest}
     \end{subfigure}
     \caption{Support for and intention of buying the 9 EUR ticket.}
     \label{fig:overall_interest}
\end{figure}

We next examine the overall support for the 9 EUR ticket as well as the intention to buy the ticket \emph{before} the introduction of the ticket on June 1, 2022. As shown in Figure \ref{fig:support}, the general support for the ticket is high in both, the Munich-oriented and the nation-wide sample. However, there is a substantial difference between the two samples: the nation-wide sample is less supportive of the ticket than the Munich-oriented sample. Given this high support for the 9 EUR ticket, we expect an equally high willingness to purchase the ticket. As the ticket is available for three months, we define a score: a person has \textit{no interest} at all if no buying intention exists for all three months, while a person has \textit{high interest} when aiming at buying the ticket in every month. Surprisingly though, the expectation of having similarly high interest compared to the ticket's support can only be partly fulfilled: looking at Figure \ref{fig:interest}, we see that the willingness to buy the ticket for the entire three months is very high in the Munich-oriented sample, while a substantial share of the nation-wide sample is not interested in buying the 9 EUR ticket at all. We find for both samples that respondents tend to either have no interest in the ticket at all or intend to buy it for the entire period in which it is offered. We further see in Figure \ref{fig:interest} that more people in the nation-wide sample compared to the Munich-wide sample do not yet consider to buy the ticket for all three months, but rather for a shorter time period; in other words, people in the nation-wide sample could be more indecisive about the ticket. Consequently, we conclude that the Munich-oriented sample has higher levels of support for and interest in buying the 9 EUR ticket, while at the nation-wide level the general high levels of support for the 9 EUR ticket do not translate directly into a buying intention for the entire period. One possible reason for this result could be that a large proportion of people in the nation-wide panel support the ticket in theory, but do not have direct access to public transport and therefore do not want to buy the ticket.  

\emph{Influence of personal opinions}

Among others, we asked participants in the questing about their opinion about phasing out coal power plants, enforcing a carbon tax, whether they prefer combating climate change over economic growth, and whether they prefer raising more tax revenue for social security schemes over lower taxes. We find higher levels of support and higher levels of buying intention for those participants that generally favor the welfare state and wish to put a stronger emphasis on combating climate change than economic growth. The latter is also reflected in a higher support and buying intention levels for the ticket when supporting an enforcement of a carbon tax and fully agreeing with a coal power plant phase out.

\emph{Influence of place of residence} 

\begin{figure}
  \centering
         \includegraphics[width=0.5\textwidth]{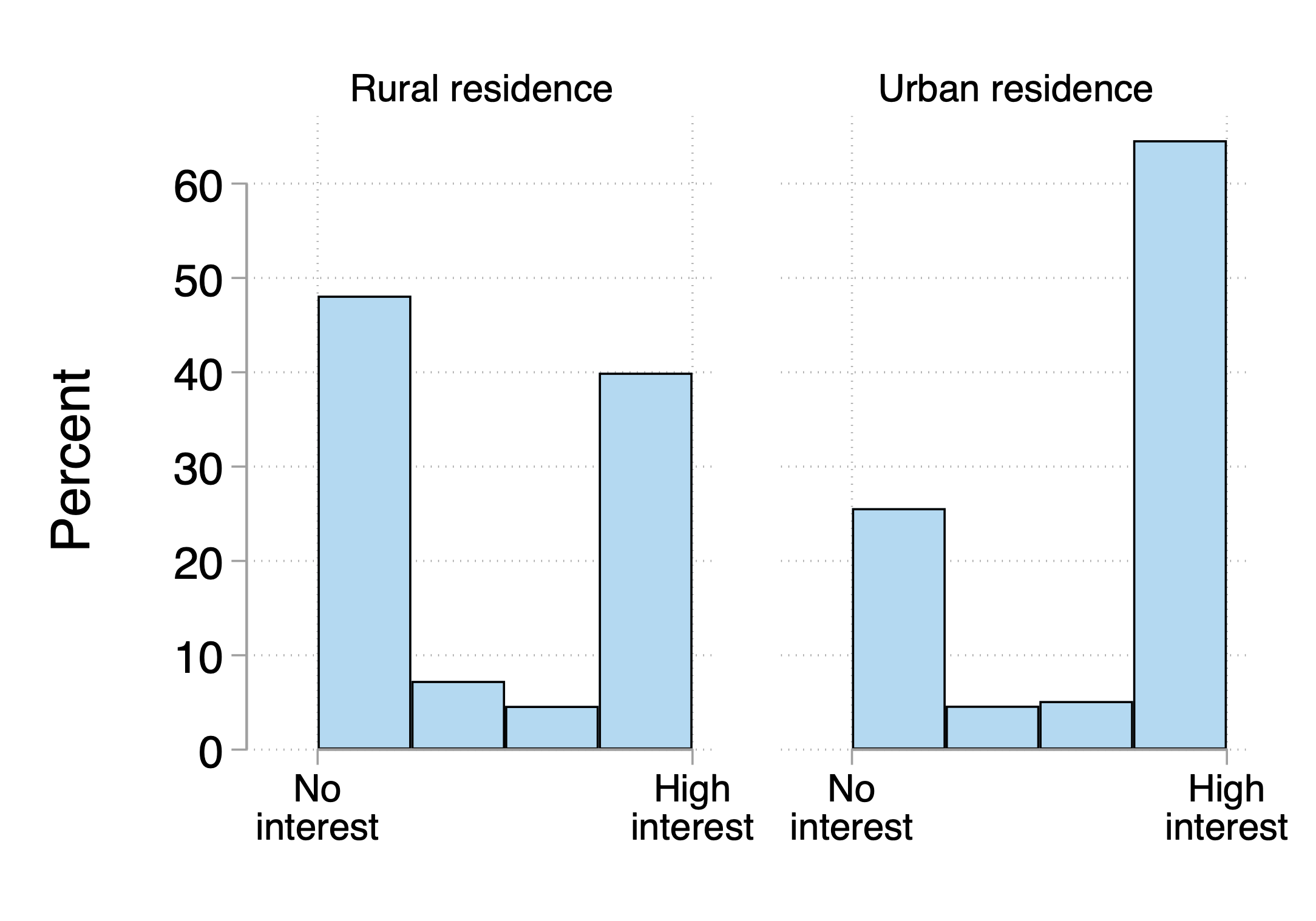}
         \caption{Intention of buying the 9 EUR ticket by residence of the pooled sample.}
         \label{fig:residence}
\end{figure}

Whether a person uses public transport largely depends on whether such services are available. Therefore, we expect that also the intention to buy the 9 EUR ticket depends on the place of residence, which we consider as each individual's primary trip origin. We distinguish between rural and urban areas according to the German RegioStar7 classification. The following results are based on the pooled sample, i.e., we do not distinguish between the Munich-oriented and the nation-wide sample. Looking at Figure \ref{fig:residence}, we find that also respondents living in rural areas show interest in buying the ticket. Though, the interest from individuals living in urban areas is much higher. This finding is line with the literature on travel behavior \cite{ewing_travel_2010}.

\emph{Influence of car ownership} 

\begin{figure}
      \centering
         \includegraphics[width=0.5\textwidth]{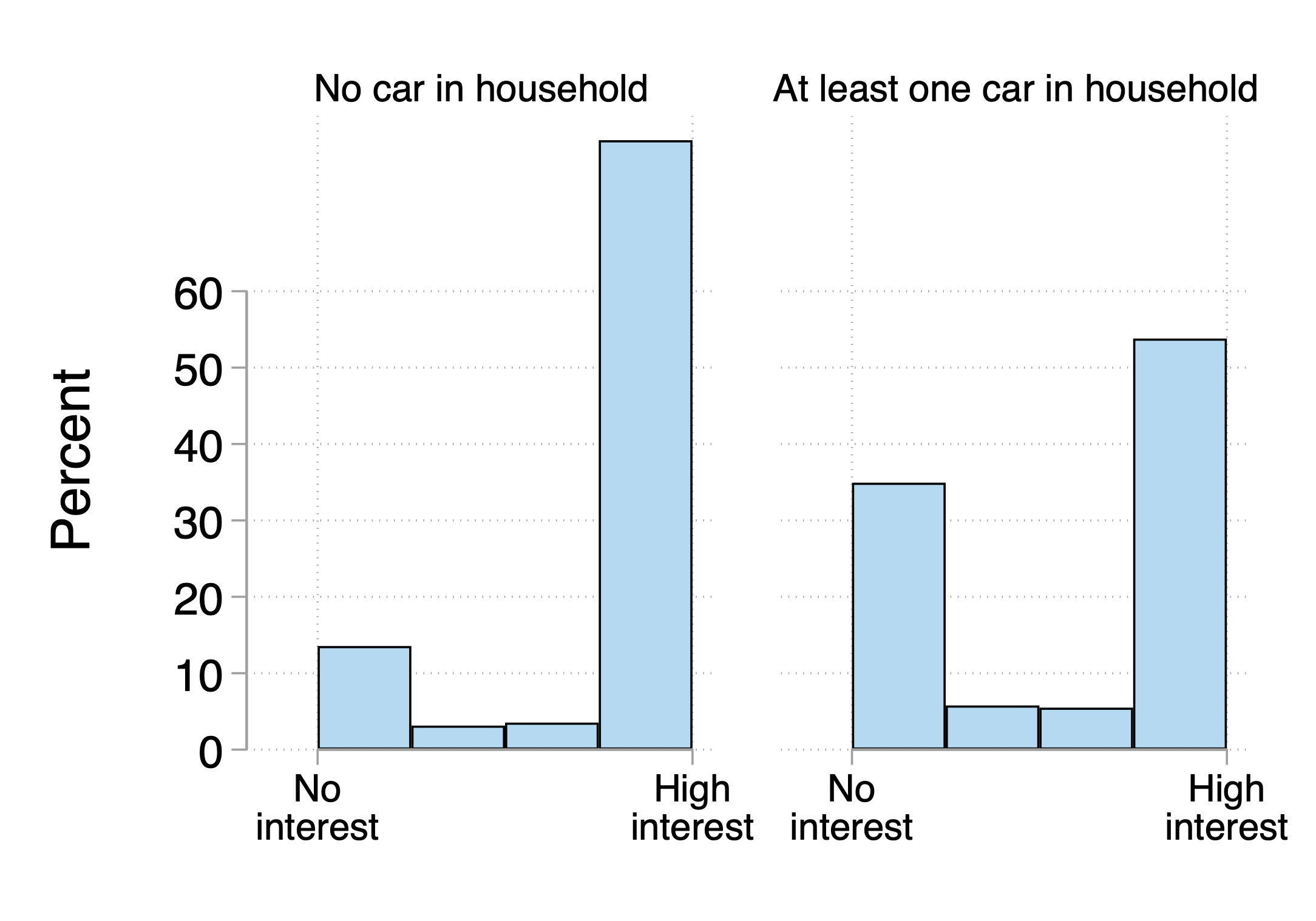}
         \caption{Intention of buying the 9 EUR ticket by car ownership of the pooled sample.}
         \label{fig:car}
\end{figure}

Another important driver of interest in the 9 EUR Ticket is the ownership of a car. We expect that car-free households are more likely to buy the 9 EUR ticket as they do not have no car alternative for mobility. Figure \ref{fig:car} shows the intention to buy the 9 EUR ticket dependent on car ownership. It is not surprising that people with no car available show high interest in the ticket. Interestingly though, we find that even if there is at least one car in the household, a significant share of the respondents show high interest in buying the ticket, too. Nevertheless, the buying intention is larger for car-free households.

\emph{Influence of household income} 

\begin{figure}
    \centering
         \includegraphics[width=0.5\textwidth]{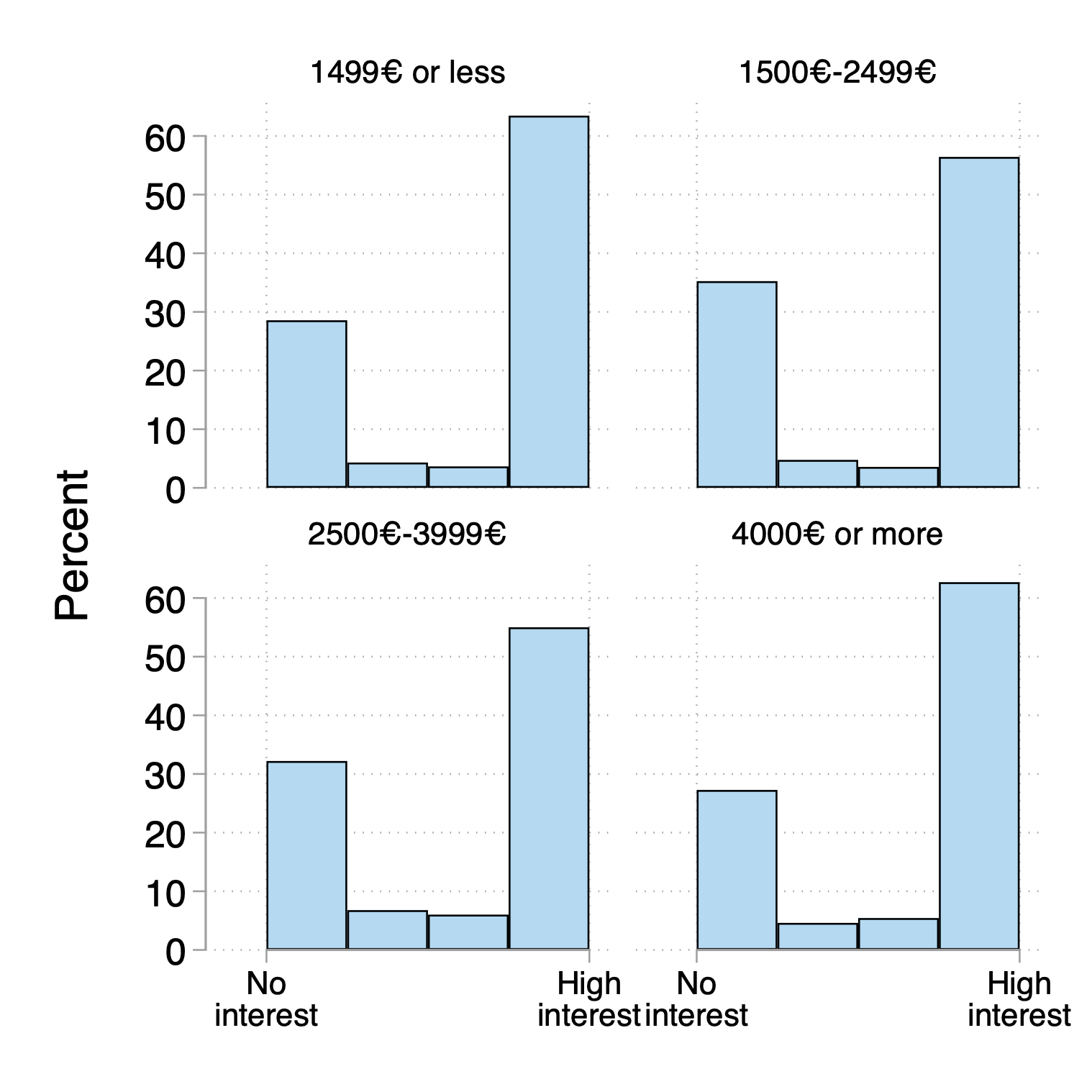}
         \caption{Intention of buying ticket by income.}
         \label{fig:income}
\end{figure}

Finally, given the recent increase in energy prices, we also expect that household income is a contributor to the general support and to the buying intention of the 9 EUR ticket. Regarding the support, we find in the pooled sample that 80\,\% in the lowest income group (1\ 500\ EUR or less) generally support the ticket, while only 5\,\% do not support the ticket. With increasing household income, support declines to approximately 70\,\% in the highest income group (4\ 000\ EUR and more), while no support increases towards 15\,\%. For the buying intention for the entire three-month period, we perhaps surprisingly do not find any pattern between income groups as seen in Figure \ref{fig:income}: Respondents in the lowest as well as in the highest income category are almost equally interested in buying the ticket. However, when we focus only on the buying intention for the month of June we find a the lowest income group (1\ 500\ EUR and less) is having a substantially higher intention or did already bought the ticket than the other income groups, which again do not exhibit any substantial variation across groups.

Interestingly, we did not find so far in the first wave evidence that higher energy costs for heating or mobility alone lead to differences in the the overall support and buying interest for the 9 EUR ticket. However, arguably it is not the total amount that matters, but the rather degree to which the recent cost increases put pressure on the budgets of households. Here we find that households that report that the cost increases are a great burden show a ten percentage point higher support level for the ticket than households where the cost increases are no burden (60\ \%). We further asked a cross-validation question about how the household deals with the price increase. Here, households who report that they are dealing badly with the situation show an five percentage point higher support level than the average household (68\ \%). Nevertheless, these two reported differences is not reflected in substantially different buying intention levels between these household types. 

Households who already save on errands do not exhibit differences in support for and intention of buying the ticket compared to households who do not; households who postpone long-term investments show a few percentage points higher support levels for the 9 EUR ticket compared to households who do not, which is also reflected in higher buying intention levels. Households in the nation-wide sample who save on leisure activities show marginally higher support levels for the 9 EUR ticket compared to households who do not; while no differences in the Munich-oriented sample are found. Importantly, these type of households show around four percentage points levels of "no intention" of buying the ticket compared to households who do not save. 

\emph{Planned analyses} 

This first look at our results includes only a small portion of the survey. In further analyses, we will not only take a closer look at other important control variables (e.g., trust in the government, political orientation), but also perform heterogeneity analyses to get a better understanding of how population groups evaluate the 9 EUR ticket. In addition, we will test different regression models to analyze the relationship between attitudes, socio-economic characteristics, stated travel behavior and the dependent variable of buying the 9 EUR ticket. Based on the data from this first wave, we can perform within-subject pre-post comparisons of the most interesting variables after collecting waves two and three later in 2022. This, in turn, allows us to draw conclusions about the change in perceptions, public opinion and travel behavior in our sample.